\documentclass{aa}  

\usepackage{physics}
\usepackage{bm}
\usepackage[dvipsnames]{xcolor}
\usepackage{graphicx}
\usepackage{hyperref}
\usepackage{bbm}
\usepackage{subcaption}
\usepackage{amsmath}
\usepackage{mathtools}
\usepackage{comment}
\usepackage{xcolor}
\usepackage{placeins}
\usepackage{textcomp}
\usepackage{txfonts}
\usepackage{soul}
\usepackage[normalem]{ulem}
\usepackage{array}
\usepackage{threeparttable}
\begin{document}

   \title{Galaxy cluster mass bias from projected mass maps}

   \subtitle{\textsc{The Three Hundred}-NIKA2 LPSZ twin samples}

   \author{M.~Mu\~noz-Echeverr\'ia\inst{\ref{LPSC}}\fnmsep\thanks{miren.munoz@lpsc.in2p3.fr}
          \and
          J.~F. Mac\'ias-P\'erez\inst{\ref{LPSC}}
          \and
          E.~Artis\inst{\ref{MPI}}
          \and
          W.~Cui\inst{\ref{Madrid}, \ref{CIAFF}, \ref{Edinburgh}}
          \and
          D.~de~Andres\inst{\ref{Madrid}}
          \and
          F.~De~Luca\inst{\ref{Roma2}}
          \and
          M.~De~Petris \inst{\ref{Roma}}
          \and
          A.~Ferragamo \inst{\ref{Roma}}
          \and
          C.~Giocoli\inst{\ref{bologna1}, \ref{bologna2}}           
          \and
          C.~Hanser\inst{\ref{LPSC}}
          \and   
          F.~Mayet\inst{\ref{LPSC}}
          \and
          M.~Meneghetti\inst{\ref{bologna1}, \ref{bologna2}}                 
          \and
          A.~Moyer-Anin\inst{\ref{LPSC}}
          \and
          A.~Paliwal \inst{\ref{Roma}}
          \and
          L.~Perotto\inst{\ref{LPSC}}
          \and
          E.~Rasia\inst{\ref{trieste1},\ref{trieste2}}
          \and
          G.~Yepes\inst{\ref{Madrid},  \ref{CIAFF}}         
     }

   \institute{
     Univ. Grenoble Alpes, CNRS, LPSC-IN2P3, 53, avenue des Martyrs, 38000 Grenoble, France \label{LPSC}
     \and
     Max Planck Institute for Extraterrestrial Physics, Giessenbachstrasse 1, 85748 Garching, Germany\label{MPI}
     \and
     Departamento de F\'isica Te\'orica and CIAFF, Facultad de Ciencias, Modulo 8, Universidad Aut\'onoma de Madrid, 28049 Madrid, Spain\label{Madrid}
     \and
     Centro de Investigaci\'{o}n Avanzada en F\'{i}sica Fundamental (CIAFF), Universidad Aut\'{o}noma de Madrid, Cantoblanco, 28049 Madrid, Spain\label{CIAFF}
     \and
     Institute for Astronomy, University of Edinburgh, Royal Observatory, Edinburgh EH9 3HJ, United Kingdom\label{Edinburgh}
     \and
     Dipartimento di Fisica, Università di Roma `Tor Vergata', Via della Ricerca Scientifica 1, I-00133 Roma, Italy\label{Roma2}
    \and
    Dipartimento di Fisica, Sapienza Universit\`a di Roma, Piazzale Aldo Moro 5, I-00185 Roma, Italy \label{Roma}
     \and
     INAF-Osservatorio di Astrofisica e Scienza dello Spazio di Bologna, Via Piero Gobetti 93/3, 40129 Bologna, Italy\label{bologna1}
     \and
     INFN-Sezione di Bologna, Viale Berti Pichat 6/2, 40127 Bologna, Italy\label{bologna2}
     \and
     INAF-Osservatorio Astronomico di Trieste, Via G. B. Tiepolo 11, 34143 Trieste, Italy\label{trieste1}
     \and
     IFPU, Institute for Fundamental Physics of the Universe, via Beirut 2, 34151 Trieste, Italy\label{trieste2}
   }

   \date{Received ...; accepted ...}

 
  \abstract{The determination of the mass of galaxy clusters from observations is subject to systematic uncertainties. Beyond the errors due to instrumental and observational systematic effects, in this work we investigate the bias introduced by modelling assumptions. In particular, we consider the reconstruction of the mass of galaxy clusters from convergence maps employing spherical mass density models. We made use of \textsc{The Three Hundred} simulations, selecting clusters in the same redshift and mass range as the NIKA2 Sunyaev-Zel’dovich Large Programme sample: $3 \leq M_{500}/ 10^{14} \mathrm{M}_{\odot} \leq 10$ and  $0.5 \leq z \leq 0.9$.  We studied different modelling and intrinsic uncertainties that should be accounted for when using the single cluster mass estimates for scaling relations. We confirm that the orientation of clusters and the radial ranges considered for the fit have an important impact on the mass bias. The effect of the projection adds uncertainties to the order of 10\% to 16\% to the mass estimates. We also find that the scatter from cluster to cluster in the mass bias when using spherical mass models is less than 9\% of the true mass of the clusters.}


   \maketitle
%

\section{Introduction}
\label{sec:intro}
    The distribution of galaxy clusters in mass and redshift is a key tool for estimating cosmological parameters \citep{Vikhlinin2009, planck2014a, Costanzi2019}. Nevertheless, the mass of clusters is not an observable quantity and has to be estimated under several hypotheses from observations. 

    Firstly, some methods assume that galaxy clusters are in hydrostatic equilibrium (HSE) and combine intra-cluster medium (ICM) observables, such as the X-ray emission and/or the Sunyaev-Zel'dovich (SZ) effect \citep{sunyaev,planck2011}, to infer the HSE mass \citep{mroczkowski2009, adam2, ruppin1, keruzore, eckert2022, munoz2022}. Secondly, others relate the velocity dispersion of the galaxies in the cluster to its total mass \citep{aguado2021,biviano2003}. Thirdly, the lensing effect on background sources due to the gravitational field of the cluster can be used to reconstruct the mass \citep{merten2015, zitrin3}. Each method is known to bias the results differently \citep{pratt2019} due to instrumental limits or departures from the assumed hypotheses, and comparison works making use of different observational results try to understand those issues \citep[e.g.][]{serenoettoricomalit}. 

    However, other biases are intrinsic to cluster physics and geometry and need to be quantified from simulations. Simulations allow one to compare the three-dimensional properties of clusters, needed for cosmological studies, to those inferred from projected maps.

    The orientation of clusters is known to affect the mass reconstruction. As shown in \citet{oguri2009}, \citet{oguriblanford2009}, \citet{hennawi2007}, \citet{ meneghetti2010a}, \citet{gralla2011}, and \cite{giocoli2023}, when clusters are elongated along the line of sight, masses are overestimated. The contrary happens when the major axes of clusters are on the plane of the sky. Thus, the projections through which clusters are observed impact their mass estimates. In \cite{meneghetti2014}, the authors explored the impact of projection effects by using the \texttt{MUSIC-2} simulations \citep{sembolini2013, biffi}, that are N-body MultiDark simulations\footnote{\url{www.MultiDark.org}} where baryons were added using smoothed-particle hydrodynamics (SPH) techniques on the dark matter, in the non-radiative flavour and without accounting for the energy feedback from Active Galactive
Nuclei (AGN). The authors estimated the mass by fitting different density models to the spherical mass density profiles of the clusters as well as to the projected mass maps. Overall, masses reconstructed from projected maps are $\sim 13-14\%$ more scattered than those estimated from the fit of three-dimensional density profiles and, according to this work, masses estimated from projected data are on average under-estimated by 5\%. They affirmed, as in \cite{giocoli2012}, that this bias is due to the orientation of clusters, which would be preferably elongated on the plane of the sky. Lensing observations give an estimate of the projected matter density distribution; therefore, the difficulty resides in recovering precise three-dimensional profiles. 

    The prolateness was also presented in \cite{giocoli2012} as a source of bias. According to simulations in the cold dark matter (CDM) framework \citep{shaw}, clusters are more frequently prolate systems than oblate-shaped. In addition, the presence of substructures was found in \cite{giocoli2012} as the second contributor to the mass bias, after halo triaxiality.

    The model chosen for the mass reconstruction can also be a source of bias. In \cite{meneghetti2014}, the authors used the Navarro-Frenk-White \citep[NFW,][]{navarro}, generalised Navarro-Frenk-White \citep[gNFW,][]{nagai, zhao}, and Einasto \citep{einasto} profile models to fit mass density radial profiles in ranges between $0.02R_{vir}$ and $R_{200}$. We note that $R_{vir}$ is the virial radius of the cluster \citep{pleebes, bryan1998} and $R_{200}$ is the radius at which the mean density of the cluster is 200 times the critical density of the Universe\footnote{\label{rdelta}The $M_{\Delta}$ and $R_{\Delta}$ are the mass and the radius where the mean density of the cluster satisfies the following: 
    $M_{\Delta}(<R_{\Delta})/\frac{4}{3}\pi R_{\Delta}^3 = \Delta \rho_{\mathrm{crit}}$. We note that $\rho_{\mathrm{crit}}$ is the critical density of the Universe at the cluster redshift, $\rho_{\mathrm{crit}} = 3 H (z)^{2}/8 \pi G$, with $H (z)$ being the Hubble function. }.
  
    They concluded that, as expected, models with three parameters (gNFW and Einasto) fit the density profile better than those with two (NFW). An important projection effect was also found regarding modelling: not all clusters that have density profiles following a NFW shape in three dimensions have a NFW-like density when projected \citep{meneghetti2014}. In \citet{giocoli2023}, the authors created \textit{Euclid}-like weak lensing observables from \textsc{The Three Hundred} simulation shear maps (see Sect.~\ref{sec:the300} for more details on the simulations). By fitting smoothly truncated NFW (tNFW) density models \citep{tnfw} to these data, they concluded that the chosen truncation radius in the model impacts the bias of reconstructed masses. In addition, they showed that the bias at $R_{200}$ and its relative uncertainty are smaller if the concentration parameter is fixed, namely at $c_{200}=3$.

    Moreover, as presented in \cite{rasia2012}, another reason to explain the bias is the approach used to select the sample. For clusters selected according to their X-ray luminosity, the reconstructed concentration-mass relation has a larger normalisation and steeper slope than the genuine relation \citep{meneghetti2014, rasia2013}. The reason behind this is that, for a given mass, the most luminous clusters are the most concentrated ones. On the contrary, if clusters are chosen for their strong lensing signal, they are preferentially elongated along the line of sight \citep{meneghetti2010a, giocoli2014} and, as a consequence, the masses obtained from projected maps are overestimated.

    Many are the effects at the origin of the bias and the scatter of the weak lensing masses of clusters. As summarised in \citet{lee2022} and references therein, assuming a NFW profile and a concentration-mass relation can introduce uncertainties in the mass from 10 to 50\% due to non-sphericity of clusters, structures along the line of sight, miscentring and halo concentration. Regarding the concentration-mass relation, from an analysis of mock galaxy cluster lenses created with the MOKA\footnote{\url{https://cgiocoli.wordpress.com/research-interests/moka/}} code \citep{giocoli2012a}, \cite{giocoli2012} concluded that the amplitude of the relation is lower when derived from lensing analyses, therefore, from projected mass tracers, than from the three-dimensional studies.
  
    In \citet{rasia2012}, 20 haloes simulated to mimic Subaru observations \citep{fabjan} were used with three projections per cluster. By fitting a NFW model to the reduced tangential shear profiles they concluded that the masses at $R_{500}^{\ref{rdelta}}$ are biased low by $\sim 7-10 \%$ with a scatter of $20 \%$. The lensing analysis based on simulations in \cite{Becker_2011} also concluded that, including shape noise only, these mass estimates are biased by $\sim 5-10 \%$ with a 20 to $30 \%$ scatter. In the same line, the weak lensing analysis in \citet{giocoli2023} derived an average $5\%$ mass bias at $R_{200}$, with the bias for a given cluster differing by up to $30\%$ depending on the orientation of the projection of the cluster observation.

    \cite{giocoli2012} investigated also the dependence of the bias with redshift and found very little evolution. Quite the opposite, they showed that the radial range chosen to fit the density or mass profile model has an important impact on the bias. According to \citet{giocoli2023}, the underestimation of cluster masses in \textit{Euclid}-like weak lensing reconstructions is more important at higher redshift, modulated by the number density of background sources.

    Some works compared also the uncertainties of masses reconstructed from total matter observables to those estimated from gas observables. They showed, on the one hand, the small impact of orientation in gas observables with simulated SZ and X-ray data \citep{Buote_2012} and, on the other hand, the irreducible non-sphericity of haloes that affects the intrinsic scatter of the weak lensing masses, tracers of the total matter distribution \citep{Becker_2011}. In the same line, \citet{meneghetti2010} suggested that lensing masses are three times more scattered ($\sim 17- 23\%$ of scatter) than X-ray estimates. The error budgets given in \citet{pratt2019} indicate that the assumption of spherical symmetry affects HSE masses at the level of a few percent, while lensing masses are affected about $10\%$ due to mass modelling. According to \cite{rasia2012}, weak lensing biases are at least twice more scattered than X-ray outcomes.

    In this work, we aim at quantifying and comparing the different contributions to the bias and its scatter. We perform the analysis in the framework of the NIKA2 Sunyaev-Zel’dovich Large Programme (LPSZ, Sect.~\ref{sec:lpsz}) with simulated clusters from \textsc{The Three Hundred} project \citep[Sect.~\ref{sec:the300},][]{cui2018}. Given the importance of the sample selection, we use clusters that are twins to those from the LPSZ (Sect.~\ref{sec:twinsamplesdef}), so that at least the mass and redshift distribution is representative of the real sample. We focus the work on total mass observables, but we also compare the results of gas observables of clusters. We compare all the masses at $R_{500}$ to be consistent with the LPSZ outcomes.

    In a recent analysis in \citet{giocoli2023}, the same \textsc{The Three Hundred} simulated data were used to study the bias of weak lensing mass reconstructions as part of the preparation work for the \textit{Euclid} mission \citep{laureijs2011euclid}. Given the observational framework they wanted to mimic, in \citet{giocoli2023} the simulated data was converted into a \textit{Euclid}-like weakly lensed galaxy distribution. With these galaxies, and accounting for the dispersion of the shapes of galaxies, they built the excess surface mass density profiles. The latter were fitted with the aforementioned tNFW model to reconstruct the three-dimensional mass density profiles of clusters. This paper does not aim to reproduce such observational effects and comparisons to \citet{giocoli2023} are valuable to elucidate the origin of some of the observed effects in that work. In addition, this paper differs from the work in \citet{giocoli2023} regarding the considered cluster sample and the overdensity of the reconstructed masses. While we restrict our analysis to a subsample of \textsc{The Three Hundred} clusters in the redshift range covered by the LPSZ programme and evaluate the masses at $R_{500}$, in \citet{giocoli2023} the authors studied the bias at $R_{200}$ for all \textsc{The Three Hundred} clusters at 9 different redshifts between $z=0.12$ and $z=0.98$. That being so, the complementarity to the analysis in \citet{giocoli2023} is clear.    
   
    This paper is organised as follows. In Sect.~\ref{sec:thissample} we present the framework of this work: the NIKA2 LPSZ, \textsc{The Three Hundred} project and the clusters selection. The different data used are presented in Sect.~\ref{sec:data}. In Sect.~\ref{sec:totmass} we describe the method followed to reconstruct the mass from convergence maps and the related bias. The different contributions to the scatter of the bias are discussed in Sect.~\ref{sec:uncertainties}. Finally, in Sect.~\ref{sec:gas} we compare the properties of mass biases obtained from total mass observables to gas observables used to reconstruct masses under hydrostatic equilibrium hypothesis. Conclusions are given in Sect.~\ref{sec:conclusions} together with the summary of our main results.


 \section{\textsc{The Three Hundred}-NIKA2 LPSZ samples}
 \label{sec:thissample}
\subsection{The NIKA2 Sunyaev-Zel’dovich Large Programme}
\label{sec:lpsz}
The NIKA2 Sunyaev-Zel’dovich Large Programme or LPSZ \citep{perotto2022, mayet2020} is a high angular resolution follow-up of $\sim 45$ clusters of galaxies. These clusters were detected with \textit{Planck} \citep{planck2016} or the Atacama Cosmology Telescope \citep[ACT,][]{hasselfield2013} and were chosen to cover a wide range of masses $(3 \leq M_{500}/ 10^{14} \mathrm{M}_{\odot} \leq 10)$ at intermediate to high redshift $(0.5 \leq z \leq 0.9)$. The LPSZ benefits from high angular SZ observations with the NIKA2 camera \citep{adam1, NIKA2-electronics,calvo} and X-ray data obtained with the XMM-\textit{Newton} and \textit{Chandra} satellites. 

By combining SZ and X-ray data, one of the objectives of the LPSZ is to re-estimate precisely the HSE $M_{500}$ mass of the galaxy clusters in the sample by reconstructing resolved hydrostatic mass profiles. This will allow us to build the relation between $M_{500}$ and the SZ signal for clusters at redshifts above 0.5 and to compare it to the relations measured at lower redshifts \citep[][hereafter A10]{arnaud10}.

However, the HSE hypothesis used to derive mass profiles is known to bias low the estimates \citep{pratt2019}. This bias is currently being investigated with different approaches \citep{gianfagna, salvati2019, de_Andres_2022}, since increasing the value of the HSE mass bias is considered as a possible solution to release the tension between the cosmological results obtained from cluster counts and CMB measurements \citep{planck2014b, planck2016a, salvati2018}. To correct for such bias, mass measurements that do not rely on the hydrostatic equilibrium assumption are needed and lensing masses can be used for this purpose. 

To tackle this issue for the NIKA2 LPSZ hydrostatic masses, we compared the HSE masses to lensing ones using the Cluster Lensing And Supernova survey with Hubble \citep[CLASH,][]{postman} data for several clusters \citep{munoz2022, munozproceedingclash, ferragamo}. Our analyses were based on the publicly available convergence maps reconstructed from weak and strong lensing data \citep{zitrin1}. As described in Sect.~\ref{sec:massreconstruction}, we fitted density models to those convergence maps, being able to get, from the best-fit density profiles, the lensing masses of clusters. In this work, we follow the same method on simulated clusters and 
we mainly focus our study on the biases that arise in the reconstruction of lensing masses from convergence maps.

\subsection{Twin samples}
\label{sec:twinsamples}
\subsubsection{\textsc{The Three Hundred} project}
    \label{sec:the300}
    This work is based on \textsc{The Three Hundred} project\footnote{\label{the300ref} \url{https://the300-project.org}} galaxy cluster simulations. \textsc{The Three Hundred} project consists of hydrodynamical re-simulations of the 324 Lagrangian regions centred on the most massive galaxy clusters identified in the \textit{MultiDarkPlanck2} box of side length $1 h^{-1}$~Gpc of the \textit{MultiDark}\footnote{Publicly available at the \url{https://www.cosmosim.org} database.} dark-matter-only simulation. The regions, of radius $15 \; h^{-1}$~Mpc and identified at $z=0$, contain clusters with virial masses above $1.2 \times 10^{15} \; \mathrm{M_{\odot}}$ and dark-matter particles of mass $1.5 \times 10^9 h^{-1} \mathrm{M}_{\odot}$. The simulation assumes a cosmology based on the \citet{planck2015} results: $h = 0.6777$, $ n = 0.96 $, $\sigma_{8} = 0.8228$, $\Omega_{\Lambda}= 0.692885$, $\Omega_{m} = 0.30711$, and $\Omega_{b} = 0.048206$. This is also the cosmological model assumed in the rest of this paper.

    The selected dark-matter-only volumes were mapped back to the initial conditions, dark matter particles were split into dark matter and gas particles and the regions were re-simulated with three hydrodynamical codes: \texttt{GADGET-MUSIC} \citep{sembolini2013}, \texttt{GIZMO-SIMBA} \citep{dave2019, Cui_2022}, and \texttt{GADGET-X} \citep{rasia2015}. We use the outputs coming from the latter, which contains very complete baryonic physics models \citep[Table 2 in][]{cui2018} with dark matter, gas, stellar, and black hole particles. Previous works based on this code have shown the agreement between simulations and observations regarding: gas density and entropy profiles \citep{rasia2015}, pressure profiles \citep{planelles2017}, and gas density and temperature at around $R_{500}$ \citep{truong2018}.

\subsubsection{Twin samples definitions}
\label{sec:twinsamplesdef}
    For our analysis, we used the clusters from \textsc{The Three Hundred} project simulations selected to constitute the twin samples of the LPSZ. As described in \citet{paliwal2021}, the clusters were chosen to cover the same redshift range as the LPSZ, that is, $0.5 \lesssim z \lesssim 0.9$. Amongst all the clusters in \textsc{The Three Hundred} satisfying this condition, we chose the snapshots 101 at $z=0.817$, 104 at $z=0.700$, 107 at $z=0.592$, and 110 at $z=0.490$. Three different samples were generated by matching properties of the clusters in the simulation to those known for the clusters in the LPSZ. The three twin samples are: 1)~$TS_{Tot-M}$, the total mass twin sample, in which the clusters in the simulation are selected so that the total $M_{500}$ from the simulation matches with the $M_{500}$ mass of the LPSZ clusters according to \textit{Planck} or ACT catalogues \citep{planck2016, Hilton2020}, 2)~$TS_{HSE-M}$, the hydrostatic mass twin sample, in which the hydrostatic mass of simulated clusters, $M_{500}^{\mathrm{HSE}}$, is matched with the \textit{Planck} or ACT masses and 3)~$TS_{Y}$, the twin sample based on the $Y_{500}$ parameter, the integrated SZ signal within $R_{500}$. 
    
    This selection gives three twin samples with 45 clusters \citep[Fig. 2 in][]{paliwal2021}. We present in Table~\ref{tab:twintable} the range of masses covered by the clusters in each twin sample. Some of the clusters being repeated, we have altogether 122 different objects. In this work, we use all the clusters from the three twin samples simultaneously. The effects studied in the following do not vary from one twin sample to another and, on the contrary, accounting for the 122 clusters improves significantly our statistics. 
\begin{table}
  \footnotesize
    \centering
    \begin{tabular}{lll}
      \hline
      \hline
    Twin sample & $M_{200}$ [$10^{14}$ M$_{\odot}$] &   $M_{500}$ [$10^{14}$ M$_{\odot}$]   \\\hline
    $TS_{Tot-M}$ &  4.19 - 16.38 (8.43) &  2.99 - 11.08 (5.68) \\
    $TS_{HSE-M}$  & 4.40 - 16.95 (8.39) & 3.23 - 11.73 (6.03) \\
    $TS_{Y}$ & 5.43 - 17.48 (9.06) & 3.94 - 12.64 (6.62) \\\hline
    \end{tabular}
    \caption{Range of masses covered by the clusters in each twin sample and the median value in brackets. We give the masses at overdensities of $\Delta= 200$ and $500$.}
    \label{tab:twintable}
     \vspace*{0.5cm}  
    \end{table}
\normalsize

\begin{figure*}[h]
        \centering
        \begin{minipage}[b]{0.49\textwidth}
        \includegraphics[trim={0pt 0pt 0pt 0pt},scale=0.35]{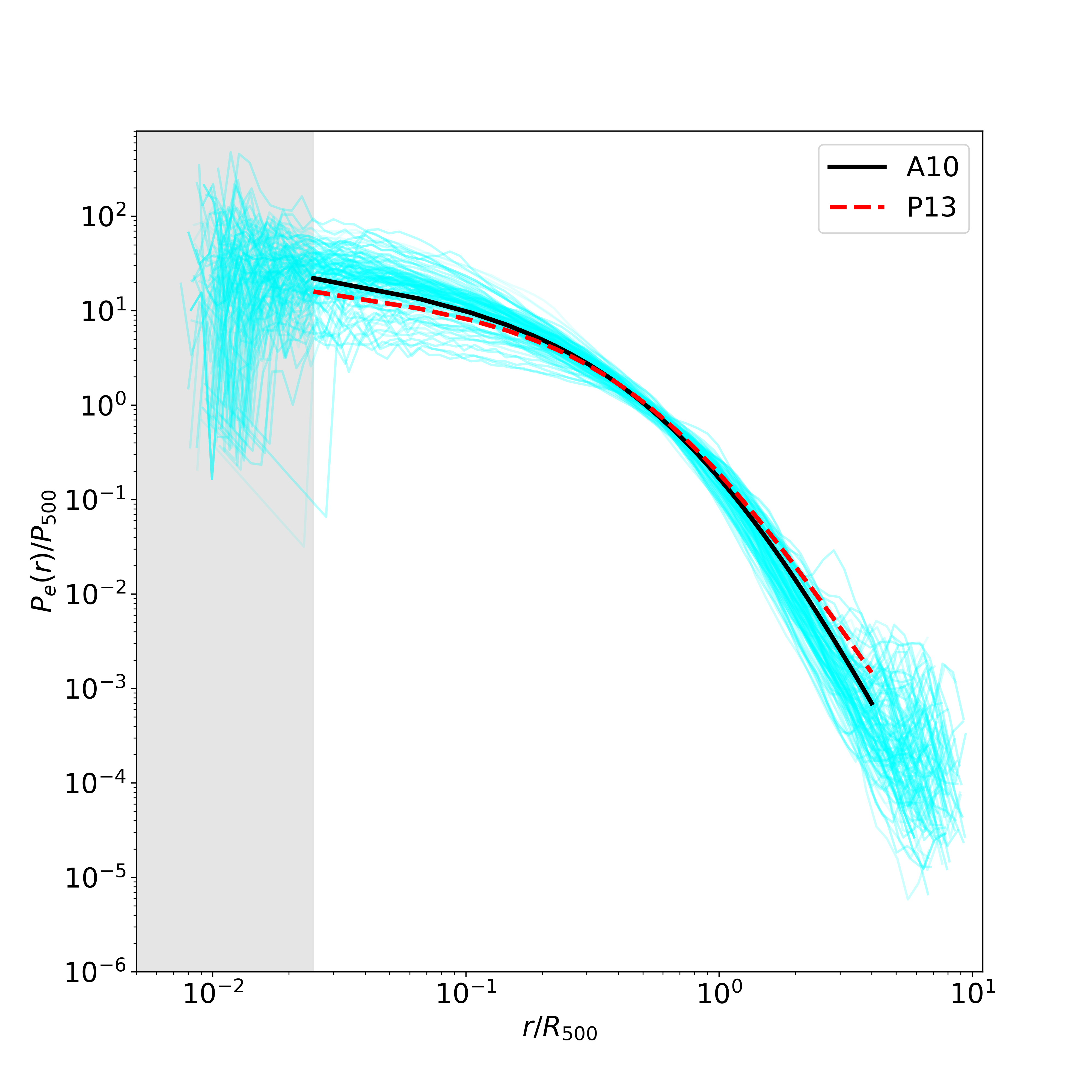}
        \end{minipage}
        \hfill
        \begin{minipage}[b]{0.49\textwidth}
        \includegraphics[trim={0pt 0pt 0pt 0pt},scale=0.35]{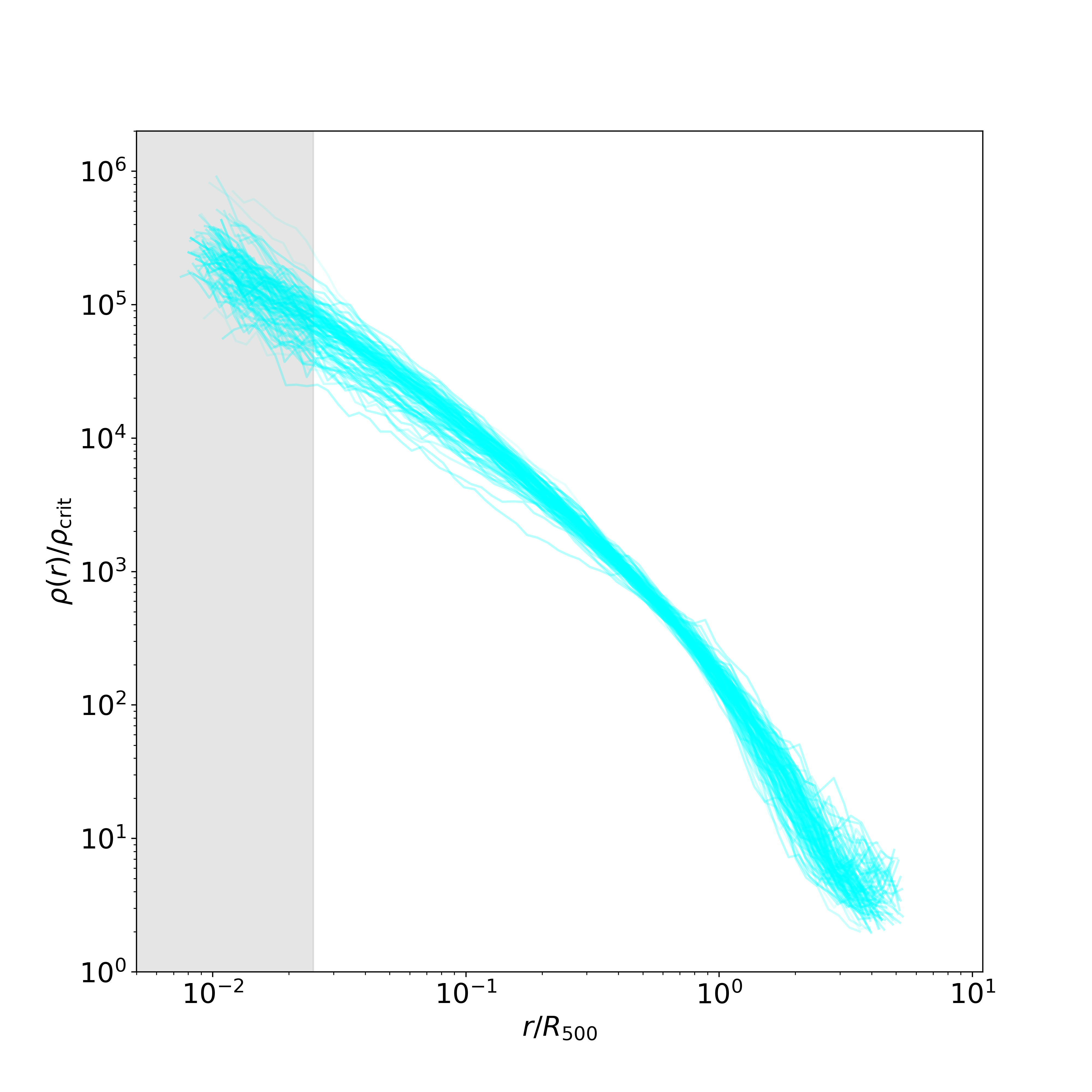}
        \end{minipage}
    \caption{Normalised gas pressure (left) and total matter mass density (right) spherical profiles for all the clusters in \textsc{The Three Hundred}-NIKA2 LPSZ twin samples. We also plot the `universal' pressure profile from the X-ray analysis in A10 and the X-ray and SZ analysis in P13. All profiles are drawn with respect to the radii normalised by the $R_{500}$ corresponding to each cluster. The grey shaded areas indicate the regions below $0.025 \times R_{500}$ that are not numerically robust.}
    \label{fig:alldensitypressure}
\end{figure*}

\section{The synthetic datasets}
\label{sec:data}
    In this section we present the data generated from the simulations and exploited in our analyses.

    \subsection{Spherical thermodynamical profiles}
    \label{sec:icmprofiles}
    For comparison sake, we make use of the spherical three-dimensional profiles of thermodynamical properties extracted from the simulations. In particular, we use the ICM pressure and total mass profiles. The profiles were computed accounting for particles in concentric shells and spheres (two-dimensional and three-dimensional profiles, respectively) centred at the maximum density peak, with radial bins starting at $10$~kpc from the centre and increasing radii by $10\%$ \citep[as in][]{Gianfagna_2022}.

    The gas pressure profiles were extracted considering hot gas particles (i.e. only particles with temperature above 0.3~keV) and following Eq.~1 in \cite{planelles}, with the correction for SPH simulations proposed in \citet{planelles} and \citet{battaglia}. The total mass profiles account for the mass of all the particles (dark matter, hot and cold gas, stars and black holes) and are used to define the true $R_{500}$ and $M_{500}$ of clusters. Profiles at $r \lesssim 0.025 \times R_{500}$ are not reliable since less than 100 particles are used to measure the thermodynamical quantities at these radial ranges.

    In Fig.~\ref{fig:alldensitypressure} we present the normalised pressure and mass density profiles for all the clusters in our sample. The pressure profiles have been normalised with respect to the $P_{500}$ obtained from Eq.~5 in A10. Following figure 3 in \cite{gianfagna}, we plot in Fig.~\ref{fig:alldensitypressure} the `universal' pressure profile from A10 as well as the profile obtained from the combination of XMM-\textit{Newton} and \textit{Planck} data in \citet{planckpressure}, hereafter P13. The density profiles were obtained by dividing the mass in each spherical shell by the volume of the shell and normalised with the critical density of the Universe at the corresponding redshift of each cluster$^{\ref{rdelta}}$. The panels in Fig.~\ref{fig:alldensitypressure} show the self-similarity of the clusters regarding the gas pressure distribution, as well as the total matter density distribution. Clusters appear consistent with the self-similar hypothesis.

    \subsection{Projected observable maps}
    \subsubsection{Projected total mass: $\kappa\text{-maps}$}
    \label{sec:kmaps}
    To study the projected total mass of galaxy clusters, we use the convergence maps  or $\kappa\text{-maps}$ produced as described in \citet{herbonnet2022} and \citet{giocoli2023}. 
    These convergence maps were generated together with shear maps to mimic the lensing effect due to the simulated clusters on background sources. The convergence of a lens at a position $\bm{\theta}$ is defined as
    \begin{equation}
        \kappa (\bm{\theta}) = \frac{\Sigma (\bm{\theta})}{\Sigma_{\mathrm{crit}}},
    \end{equation}
    where $\Sigma (\bm{\theta})$ is the projected mass density of the cluster at $\bm{\theta}$ and $\Sigma_{\mathrm{crit}}$ the so-called critical surface density:
    \begin{equation}
       \Sigma_{\mathrm{crit}} = \frac{c^2}{4\pi G}\frac{D_{S}}{D_{L}D_{LS}}.
    \end{equation}
    Here $D_{S}$, $D_{L}$, and $D_{LS}$ are the angular diameter distances between the observer and the background source, the observer and the lens (the cluster), and the source and the lens, respectively. 
    
    To create the convergence maps, first the $\Sigma\text{-maps}$ were obtained by projecting, within a volume of depth 10~Mpc, the masses of all the particles of the clusters along different axes. Then, these $\Sigma\text{-maps}$ were divided by $\Sigma_{\mathrm{crit}}$, assuming all background sources to be at $z = 3$. For each cluster, maps of $6 \times 6$~Mpc$^2$ were produced, centred on the minimums of the gravitational potential wells. To avoid boundary errors, final maps are $5 \times 5$~Mpc$^2$ with $2048 \times 2048$~pixels, which gives different angular resolution pixels for different redshift clusters. 
    
    Six different maps are available for each cluster, corresponding to different projection axes. Three maps (hereafter 0, 1, and 2) are obtained from the projection along the $x$, $y$, and $z$ orthogonal axes of the simulation. In principle, they correspond to random directions with respect to the morphology of the clusters. The other three (0\_pr\_axes, 1\_pr\_axes, and 2\_pr\_axes) are aligned with the principal axes of the clusters regarding their moments of inertia \citep{Knebe_2020}.
    
    \subsubsection{Sunyaev-Zel'dovich effect: $y\text{-maps}$}
    \label{sec:ymaps}
    The gas in the ICM of galaxy clusters 
    is investigated starting from the $y\text{-maps}$. The thermal SZ effect due to galaxy clusters is characterised by the Compton parameter, proportional to the integral along the line of sight of the thermal pressure $P_{e}$ in the ICM,
    
    \begin{equation}
	    y = \frac{\sigma_{\mathrm{T}}}{m_{e}c^{2}} \int_{0}^{+\infty} P_{e} \hspace{0.3pt} \mathrm{d}l = \frac{\sigma_{\mathrm{T}}k_{\mathrm{B}}}{m_{e}c^{2}} \int_{0}^{+\infty} n_{e}T_{e} \hspace{0.3pt} \mathrm{d}l,
	    \label{eq:compton}
    \end{equation}   
    with $\sigma_{\mathrm{T}}$, $k_{\mathrm{B}}$, $m_{e}$, and $c$ the Thomson cross section, the Boltzmann constant, the electron rest mass, and the speed of light, respectively. Assuming the ICM behaves as an ideal gas, the Compton parameter is also directly proportional to the integral of the electron number density $n_e$ and the electron temperature $T_e$. 
    
    As described in \citet{cui2018} and previously done in \citet{sembolini2013} and \citet{lebrun2015}, in \textsc{The Three Hundred} project this integration is converted into a summation\footnote{Using the publicly available Python package: \url{https://github.com/weiguangcui/pymsz.}} over the gas particles in the line of sight and within a $2\times R_{200}$ depth volume,
    \begin{equation}
        y = \frac{\sigma_{\mathrm{T}}k_{\mathrm{B}}}{m_{e}c^{2} \mathrm{d}A} \sum_{i=0}^{N_{\mathrm{gas \; part.}}} T_{e,i} N_{e,i} W(r,h_i).
   \label{eq:ysimu}
    \end{equation}
    Here the electron number density $n_e$ is represented as the number of electrons in a given gas particle, $N_e$, divided by its spatial volume $\mathrm{d}V$ ($\mathrm{d}V = \mathrm{d}l \mathrm{d}A$): $n_e =  N_e/\mathrm{d}V =  N_e/\mathrm{d}A/\mathrm{d}l$. In Eq.~\ref{eq:ysimu} $W(r,h_i)$ corresponds to the SPH smoothing kernel used to smear the signal of each particle to the projected pixels, with smoothing length $h_i$ \citep{cui2018}. 
    
     For each cluster, there are 29 $y\text{-maps}$ available, projected along the $x$, $y$, and $z$ main axes of the simulation, plus along other 26 random axes. We analysed only the projections in common with the $\kappa\text{-maps}$, hence the three main axes projections. Centred on the projected maximum density peak of the clusters, maps have $1920 \times 1920$~pixels of $5^{\prime\prime}$ angular resolution, going up to 14 to $30 \times R_{200}$.

    \subsubsection{Projected gas mass: gas mass maps}
    \label{sec:gasmassmaps}
    A different way to characterise the distribution of the gas in the ICM is to use projected gas mass maps. They were generated as the $\Sigma\text{-maps}$ (Sect.~\ref{sec:kmaps}), but accounting only for the gas particles along the line of sight. Final maps are also $5 \times 5$~Mpc$^2$ with $2048 \times 2048$~pixels and projected along the three main axes of the simulation.


\section{Total mass reconstruction from $\kappa\text{-maps}$}
\label{sec:totmass}

Just as in \cite{meneghetti2014}, in this work we do not consider any systematic effect, such as the accuracy of galaxy shape measurements or photometric redshifts, that may affect lensing analyses. Those effects can introduce additional uncertainties that require separate investigations \citep{Bah__2012, giocoli2023}. We restrict our work to the effects that arise from the reconstruction of three-dimensional quantities from projected data.
    
    \subsection{Mass reconstruction procedure}
    \label{sec:massreconstruction}
    \begin{figure*}[h]
        \centering
        \begin{minipage}[b]{0.49\textwidth}
        \includegraphics[trim={0pt 0pt 0pt 0pt},scale=0.35]{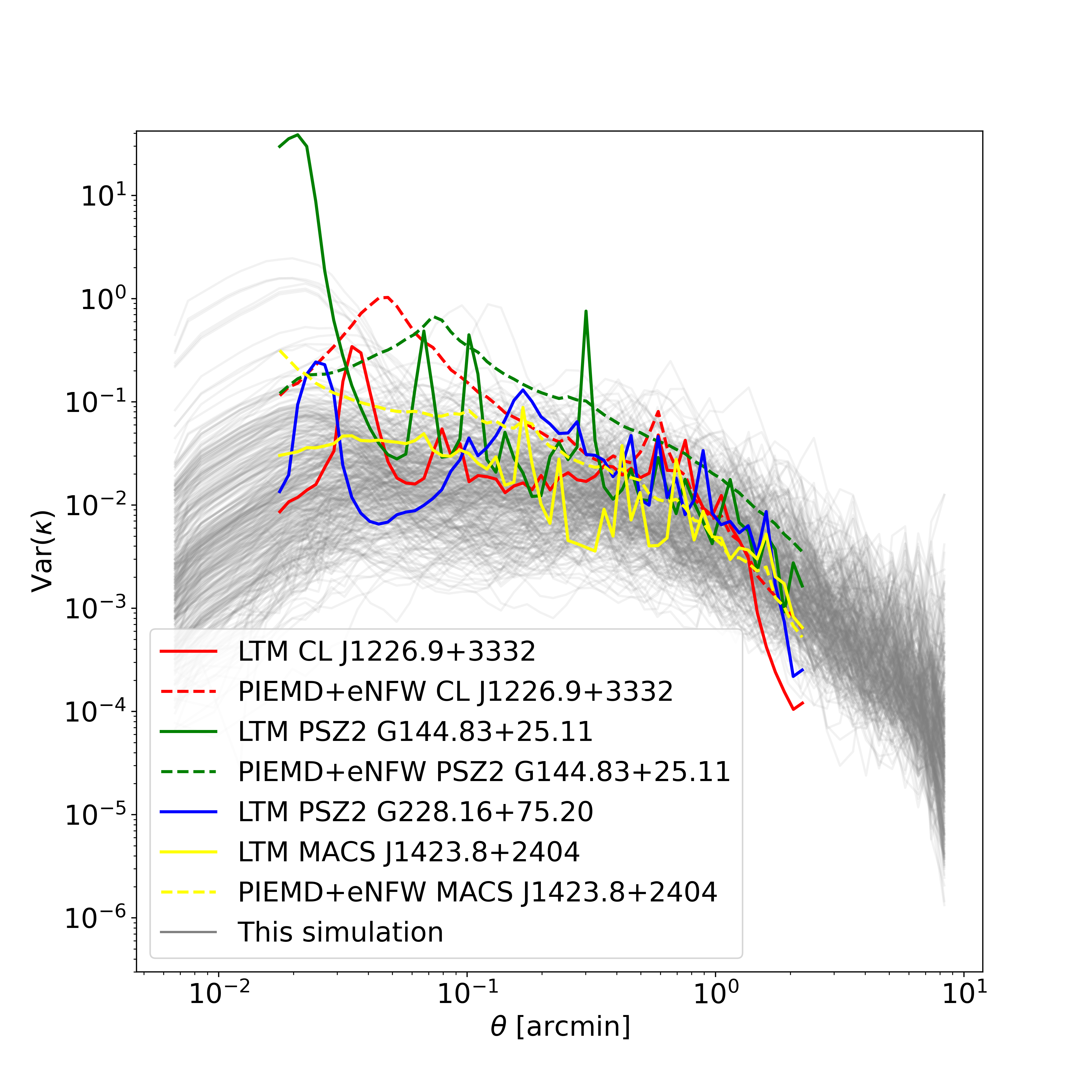}
        \end{minipage}
        \hfill
        \begin{minipage}[b]{0.49\textwidth}
        \includegraphics[trim={0pt 0pt 0pt 0pt},scale=0.35]{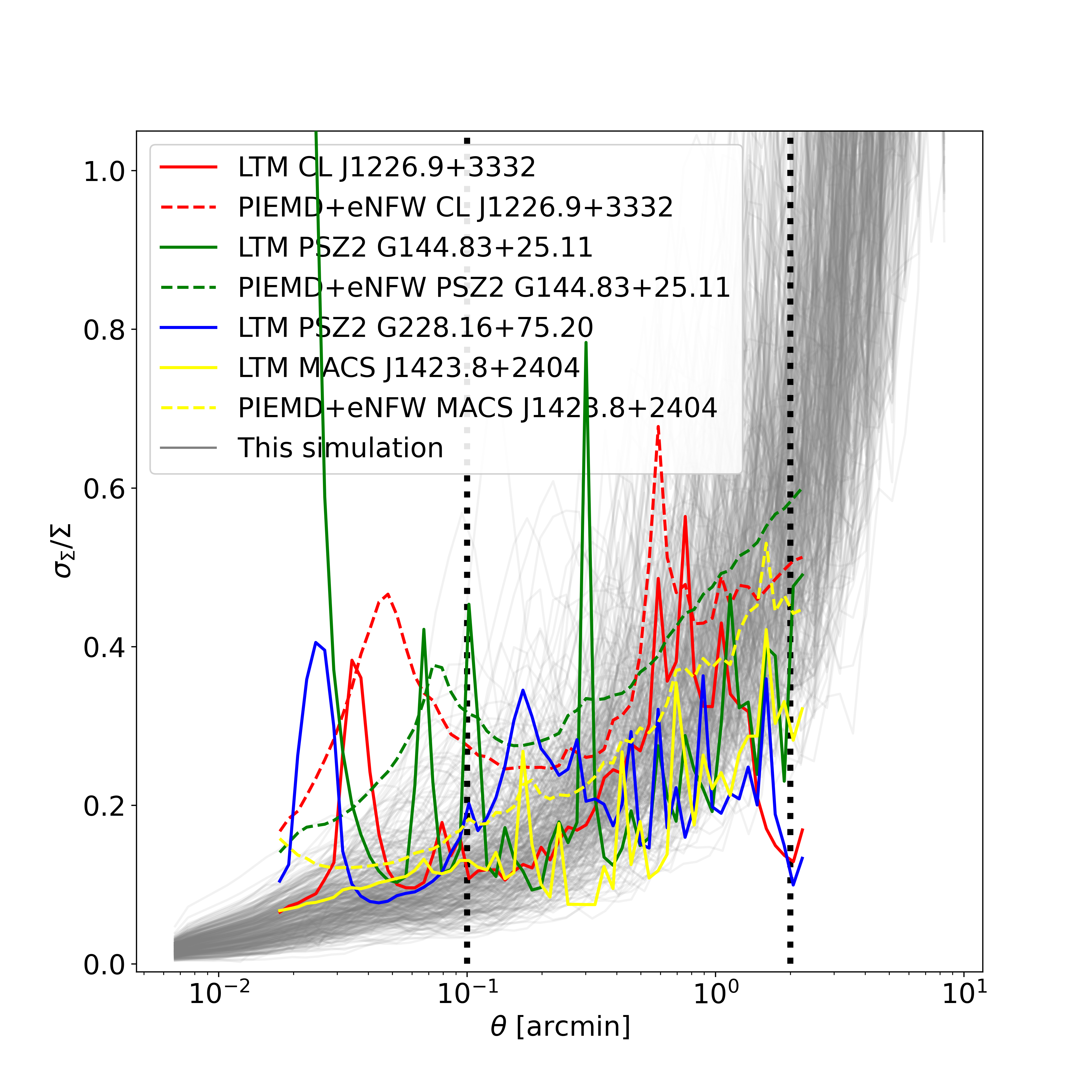}
        \end{minipage}
    \caption{Variance of the convergence profiles (left) and fractional error of the $\Sigma$-profiles (right) calculated from convergence maps. In grey we show the profiles for the simulated clusters in \textsc{The Three Hundred}-NIKA2 LPSZ samples. Red, green, blue, and yellow profiles correspond to CL J1226.9+3332, PSZ2 G144.83+25.11, PSZ2 G228.16+75.20, and MACS J1423.8+2404 clusters, respectively, with data from the CLASH convergence maps \citep{zitrin1} and uncertainties computed following \citet{munoz2022} \citep[sample defined in][]{munozproceedingclash}. Solid and dashed lines differentiate the profiles for the CLASH convergence maps reconstructed assuming the Light-Traces-Mass (LTM) and the Pseudo-Isothermal Elliptical Mass Distribution with an elliptical NFW (PIEMD+eNFW) models, respectively. Vertical dotted lines in the right panel indicate the minimum and maximum of the used radial range.}
    \label{fig:choiceofrange}
    \end{figure*}
    
\begin{figure*}
  \centering
  \includegraphics[trim={0pt 0pt 0pt 0pt},scale=0.38]{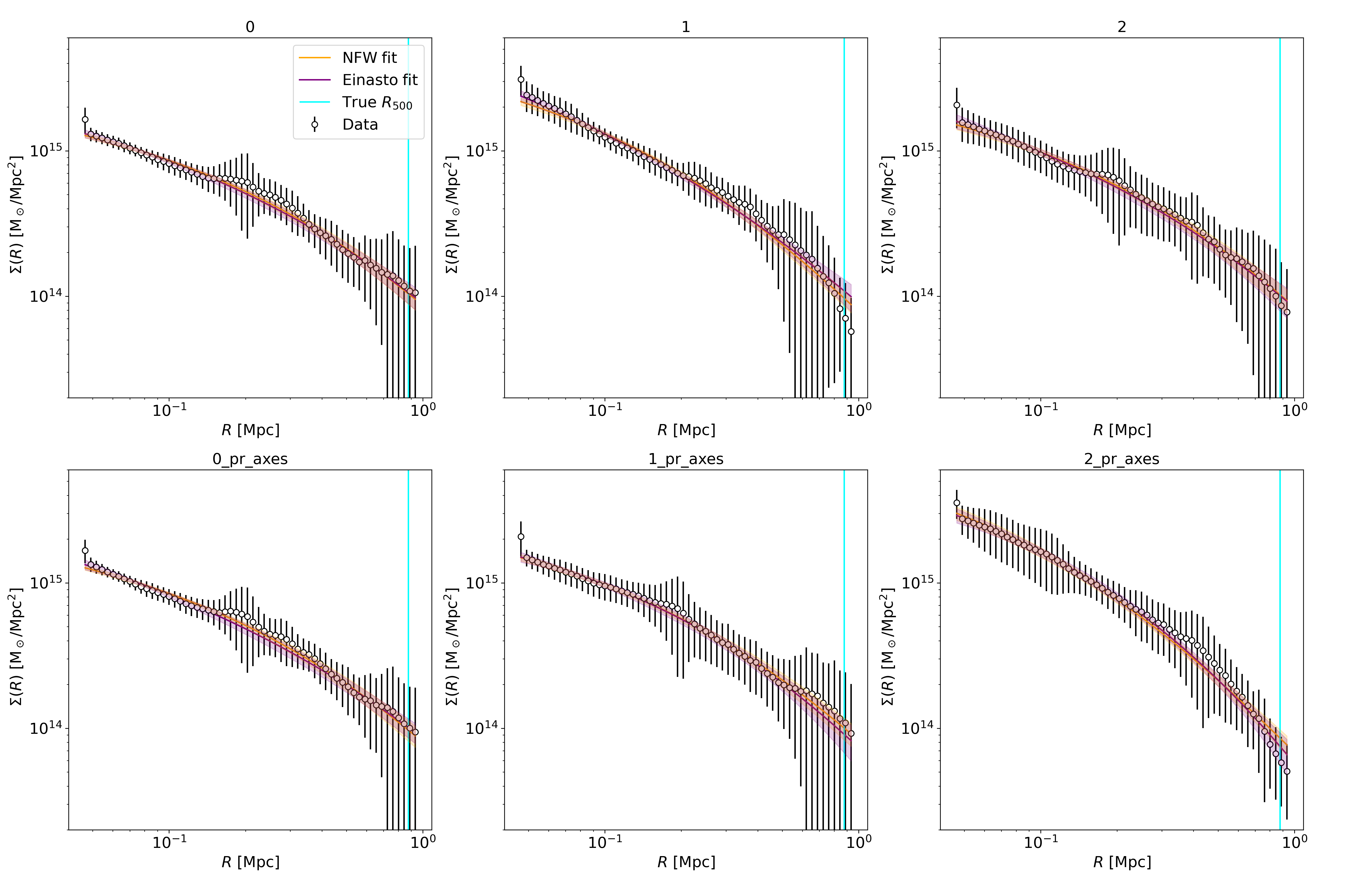}
  \caption{Projected mass density profiles (empty markers) for the six convergence maps for the 0306 cluster in the snapshot 101 ($z=0.817$). The orange and purple profiles show the fit with NFW and Einasto models, respectively. We give the mean profiles with $1\sigma$ contours. From left to right and top to bottom we present the 0, 1, and 2 random projections and 0\_pr\_axes, 1\_pr\_axes, and 2\_pr\_axes principal moment of intertia projections. The vertical cyan line shows the true $R_{500}$ of the cluster.}
  \label{fig:sigmafit}
\end{figure*}
 
    As aforementioned, to estimate the total mass of the clusters from projected maps we followed the procedure used in \citet{ferragamo}, \citet{munozproceedingclash}, and \citet{munoz2022}. In these works convergence maps from the CLASH dataset are used to reconstruct the lensing mass of galaxy clusters: $\kappa\text{-maps}$ are converted into projected mass density maps, $\Sigma$, and mass density models are fitted to the radial profile of $\Sigma$, which allows one to reconstruct the lensing mass. 
    
    From the convergence maps described in Sect.~\ref{sec:kmaps}, we obtain projected $\Sigma$-profiles by radially averaging the values in the map, starting from their central pixel. Uncertainties are computed from the dispersion in each radial bin to account for non-circular features in the map. In a previous work \cite{giocoli2012} showed the important impact of the chosen radial range to perform the fit. To build the excess surface mass density profiles in \citet{giocoli2023} the authors used 22 logarithmically spaced bins, spanning from $0.02$ to $1.7 h^{-1}$~Mpc from the cluster centre.

    We decided to consider instead a radial range in which the fractional errors of the $\Sigma$-profiles are comparable to the uncertainties obtained from data-based convergence maps. In the left panel in Fig.~\ref{fig:choiceofrange}, we show the variance of binned profiles obtained from convergence maps. In grey we present the profiles obtained from the simulated convergence maps. We aim to assess the lensing mass reconstruction of the NIKA2 LPSZ clusters \citep{ferragamo,munozproceedingclash,munoz2022} based on the CLASH convergence maps \citep{zitrin1} obtained with the highest quality lensing data. Thus, we show in red, green, blue, and yellow in Fig.~\ref{fig:choiceofrange} the variance of the convergence profiles obtained from CLASH convergence maps for the four clusters (with two convergence map models for three of them) analysed in \citet{munoz2022}, \citet{munozproceedingclash}, and \citet{ferragamo} \citep[for more details on the uncertainties of the profiles see][]{munoz2022}. We observe an overall agreement, with possible differences in the variance trends in regions $< 0.02$ arcmin, where all simulation profiles tend to rapidly decrease.

    In the right panel in Fig.~\ref{fig:choiceofrange} we compare the fractional errors of the $\Sigma$-profiles from the simulation and from CLASH maps. The agreement is again very good, but in the region between $\sim 0.02$ and $\sim 0.05$ arcmin the observed errors tend to be larger than the average simulated errors. For this reason, we restricted the fit of the $\Sigma$-profiles to the radii above $0.1^{\prime}$, removing the radial ranges where the fractional uncertainties of simulated $\Sigma$-profiles are systematically smaller than data-based error bars and, therefore, unrealistic. Regarding the upper limit, we restricted the fit of the profiles to $2^{\prime}$, which is the maximum radial extent of convergence maps from \citet{zitrin1}.

    To model the $\Sigma$-profile we used the two-parameter NFW mass density model:
    \begin{equation}
        \rho_{\mathrm{NFW}}(r) = \frac{\rho_s}{(r/r_s)(1+r/r_s)^2} =\frac{c_{\Delta}^3 \Delta \rho_{\mathrm{crit}}/3 I(c_{\Delta})}{(r c_{\Delta}/ R_{\Delta}) (1+r c_{\Delta}/ R_{\Delta})^2} , 
    \label{eq:nfw}
    \end{equation}
    and the three-parameter Einasto model,
    \begin{equation}
      \rho_{\mathrm{Ein}} (r)= \rho_s \; \mathrm{exp}\left\lbrace -\frac{2}{\alpha}\left[\left(\frac{r}{r_s}\right)^{\alpha}-1\right] \right\rbrace,
      \label{eq:einasto}
    \end{equation}
    and fitted their projected mass density profiles to the $\Sigma$-profiles from the convergence maps:
    \begin{equation}
      \Sigma_{\mathrm{model}}(R) = 2 \int_{0}^{+\infty} \rho_{\mathrm{model}} \left(\sqrt{r^2+R^2} \right)\; \mathrm{d}r.
    \end{equation}
    Here $r$ is the radius of three-dimensional spherical profiles while $R$ is used for the projected profiles. In Eq.~\ref{eq:nfw}, $c_{\Delta}$ is the concentration parameter that corresponds to $R_{\Delta}$, $c_{\Delta} = R_{\Delta} /r_{s}$, and $I(c_{\Delta})$ is a function that depends only on $c_{\Delta}$ and the shape of the density profile \citep[Eq. 15 in][]{aguena2021}, which for the NFW model gives
\begin{equation}
  I(c_{\Delta}) =  \ln (1+c_{\Delta}) -  \frac{c_{\Delta}}{1+c_{\Delta}}.
\end{equation}         
The parameters for the Einasto density model (Eq.~\ref{eq:einasto}) are $r_s$, $\rho_s$, and $\alpha$: the scale radius, the characteristic density, and the shape parameter, respectively.
    
    The fits were performed via a Markov chain Monte Carlo (MCMC) analysis using the \texttt{emcee} Python package \citep{foreman,goodman}. In addition, we made use of
    the \texttt{profiley} Python package\footnote{\url{https://profiley.readthedocs.io/en/latest/index.html}.}, which contains already tested \citep{act2020} functions that describe NFW and Einasto spherical density profiles, as well as the corresponding line of sight projected profiles. 
    Regarding the prior ranges of the model parameters, we considered uniform priors \citep[as in][]{ettori2019} wide enough to cover all possible results found in the literature. For the NFW model we took $c_{200}= \mathcal{U}(0.1, 15)$ and $r_{s} = \mathcal{U}(0.1, 6)$~Mpc, which are safe ranges in view of the concentration and mass values obtained for observational results in \citet{ettori2019} and \citet{umetsu2020}. For Einasto we considered $r_{s} = \mathcal{U}(0.05, 3)$~Mpc, $\rho_{s} = \mathcal{U}(10^{10}, 10^{18})$~M$_{\odot}$/Mpc$^{3}$, and $\alpha = \mathcal{U}(0, 5)$ to make sure to cover broadly the variety of possible results \citep{ettori2019, eckert2022-2}. Once the fit was performed, we verified that the modelled $\Sigma$-profile is a good representation of the data following a goodness of fit criterion defined as
    \begin{equation}
      \label{eq:fitquality}
      \text{Fit quality = median}( | \Sigma_{\text{data}}\; - \; \Sigma_{\text{model}} |  / \sigma_{\Sigma_{\text{data}}}  ), 
    \end{equation}
    where the median is estimated over all the radial bins and posterior distributions of the model. Bad fits were rejected if $\mathrm{Fit\; quality} > 1$. This means that if the median of the absolute difference between the $\Sigma$-profile data bins and the model with respect to the uncertainties is larger than unity, the fit is not satisfactory.

    After the fit, mass profiles were obtained by computing for each sample of parameters the corresponding profile. For NFW model, the spherical mass profile is described by, 
    \begin{equation}
    \begin{split}
        M_{\mathrm{NFW}}(<r) = 4\pi \rho_s r_{s}^{3} \left[ \frac{1}{1+r/r_{s}} + \ln (1+r/r_{s}) -1  \right] = \\
        4\pi c_{200}^3 200 \rho_{\mathrm{crit}}/3 I(c_{200}) (R_{200}/c_{200})^{3}\\
        \left[ \frac{1}{1+ r c_{200}/R_{200}} + \ln (1+rc_{200}/R_{200}) -1  \right].
    \end{split}
    \end{equation}
    And for Einasto,
    \begin{equation}
        M_{\mathrm{Ein}}(<r)=\frac{4\pi\rho_sr_{s}^3}{\alpha}e^{2/\alpha}\left(\frac{\alpha}{2}\right)^{3/\alpha}\gamma\left[\frac{3}{\alpha}, \frac{2}{\alpha}\left(\frac{r}{r_s}\right)^{\alpha}\right],
    \end{equation}
    where $\gamma$ is the incomplete lower gamma function:
    \begin{equation}
        \gamma(a,x) = \int_{0}^{x} t^{a-1}e^{-t}\mathrm{d}t.
    \end{equation}

    Finally, from the reconstructed mass profiles we computed the $M_{500}^{\kappa}$ mass of the cluster at its corresponding $R_{500}^{\kappa}$ (Fig.~\ref{fig:massfit}). This procedure was repeated for each $\kappa\text{-map}$, therefore, several times for each cluster, which gave different $M_{500}^{\kappa}$ estimates affected by the projection effect and the choice of the density model. As an example, in Fig.~\ref{fig:sigmafit} we present the $\Sigma$-profiles for the six projections of one cluster (number 0306) at $z=0.817$. The orange and purple lines show the mean profiles (solid lines) and $1\sigma$ error bars (shaded areas) obtained from the posterior distributions of NFW and Einasto fit parameters. In the 2\_pr\_axes projection the cluster is observed along its most elongated axis. For this reason, the centre appears very massive and the density in the outskirts drops fast. We verified that the sizes of the uncertainties for these $\Sigma$-profiles are compatible with the ones obtained from the convergence maps built by combining weak and strong lensing data. Nevertheless, they do not necessarily correspond to what we would expect from other observations. For instance, in weak lensing mass density reconstructions error bars tend to be larger in the centre and smaller in the outskirts of clusters.

    We repeated the analysis for all the clusters in our sample and for the following we take, amongst all the $\Sigma$-profile fits, the results that pass the quality criteria in Eq.~\ref{eq:fitquality} for both the NFW and Einasto fits. As explained in Sect.~\ref{sec:projection}, in the rest of the study we only consider the three projections along the $x, y$, and $z$ orthogonal axes of the simulation.

    We study the correlations between the best-fit values of the posterior distributions of model parameters for all the clusters (shown in Fig.~\ref{fig:fitparams} in Appendix~\ref{sec:appendixA}). An anticorrelation between Einasto’s $\rho_s$ and $r_s$ parameters is clearly noticeable, with a Pearson correlation coefficient of $-0.56$. Contrary to some studies \citep[][and references therein]{lopezcano} showing a decreasing tendency of the concentration, $c_{200}$, with the mass, $M_{200}$, we do not find any significant correlation. In \citet{cui2018} the concentration-mass relation of $z=0$ clusters from \textsc{The Three Hundred} simulation is presented, showing that the relation is flatter for \texttt{GADGET-X} clusters than for \texttt{GADGET-MUSIC}. Some works \citep[e.g.][]{oguri2009} also show a dependence of the concentration parameter with redshift, but our results do not present any clear evolution along cosmic time (the Pearson correlation coefficient between the redshift and the concentration is of r $= -0.09$). According to \cite{meneghetti2014}, the redshift evolution of the concentration-mass relation is weak for the \texttt{MUSIC-2} simulation.

    \subsection{Mass bias from convergence maps}
    We define the mass bias as the relative distance of each reconstructed mass, $M_{500}^{\kappa}$ (as above-mentioned, evaluated at the $R_{500}^{\kappa}$ obtained from the reconstructed mass profile), to the true mass of the cluster, $M_{500}$:
    \begin{equation}
      \label{eq:biask}
         b^{\kappa} = (M_{500} - M_{500}^{\kappa}) / M_{500}.
    \end{equation}     
    \begin{figure*}
        \centering
        \includegraphics[trim={0pt 0pt 0pt 0pt}, scale=0.45]{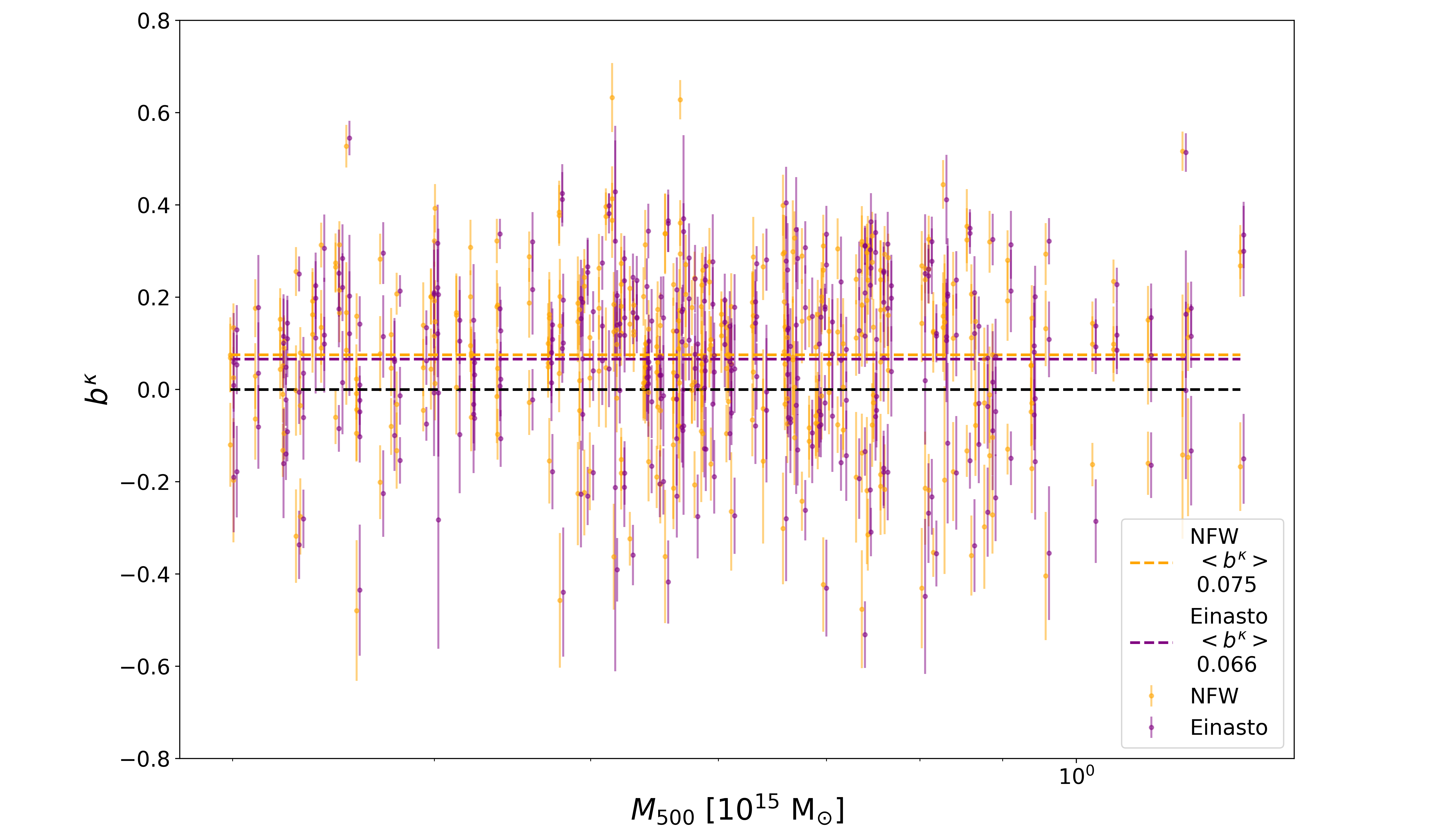}
        \caption{Mass bias for masses inferred from fitted $\kappa$-maps with respect to $M_{500}$ per cluster. Each data point corresponds to the mean bias of a cluster for a given random projection (0, 1 or 2) with error bars showing the $1\sigma$ uncertainty. Results for the Einasto model have been artificially shifted in mass for visualisation purposes. The horizontal orange and purple dashed lines show the mean for all the results for NFW and Einasto, respectively. The black line indicates the zero.}
        \label{fig:kappabias}
    \end{figure*}            
    \begin{figure}
      \centering
      \includegraphics[scale=0.42]{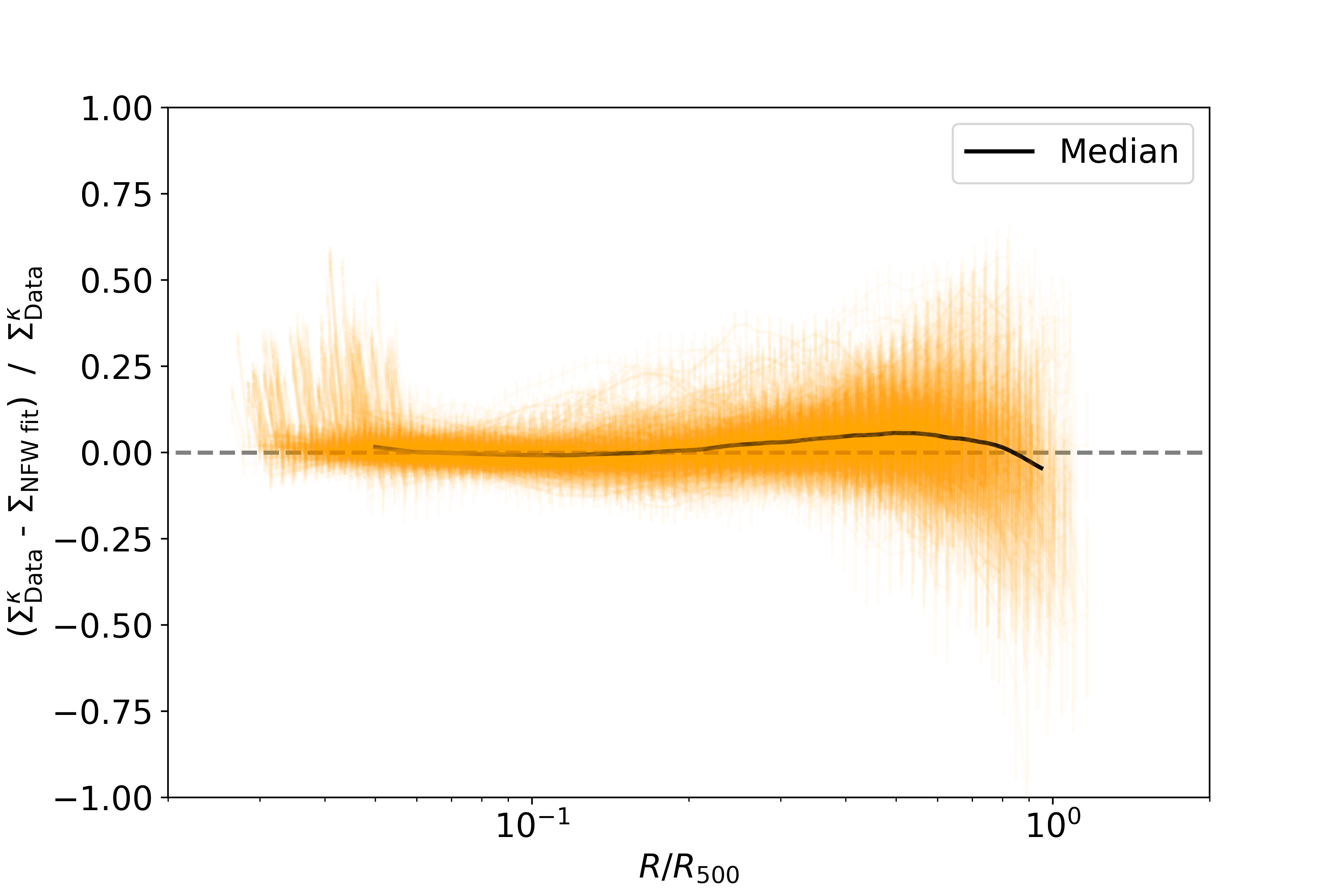}
      \includegraphics[scale=0.42]{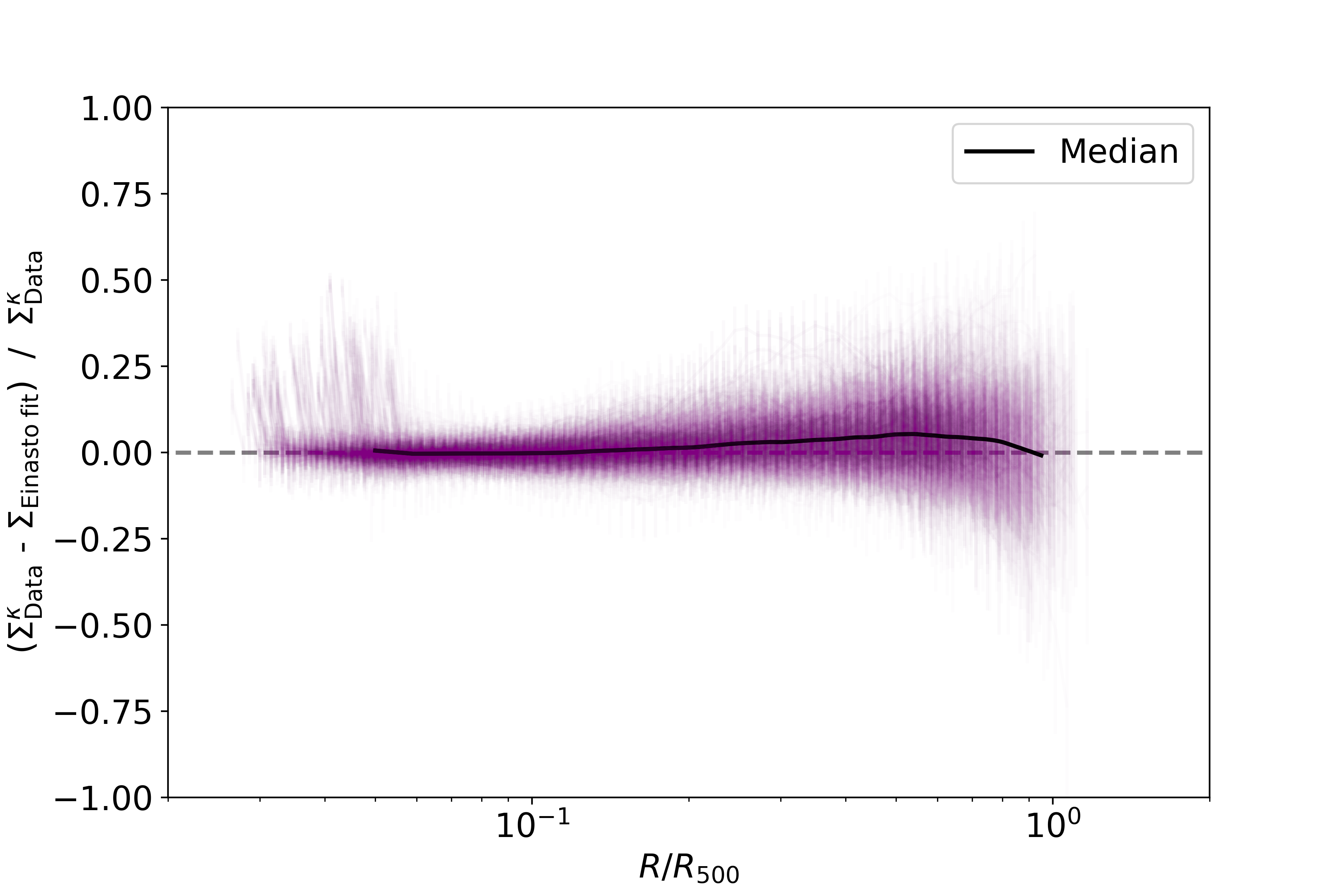}  
    \caption{Relative differences of $\Sigma$-profiles between the fits with NFW (top) and Einasto (bottom) models and the profiles measured from the convergence maps. Solid lines show the median.}
    \label{fig:nfweinastofits}
    \end{figure}
        \begin{figure}
        \centering
        \includegraphics[scale=0.33]{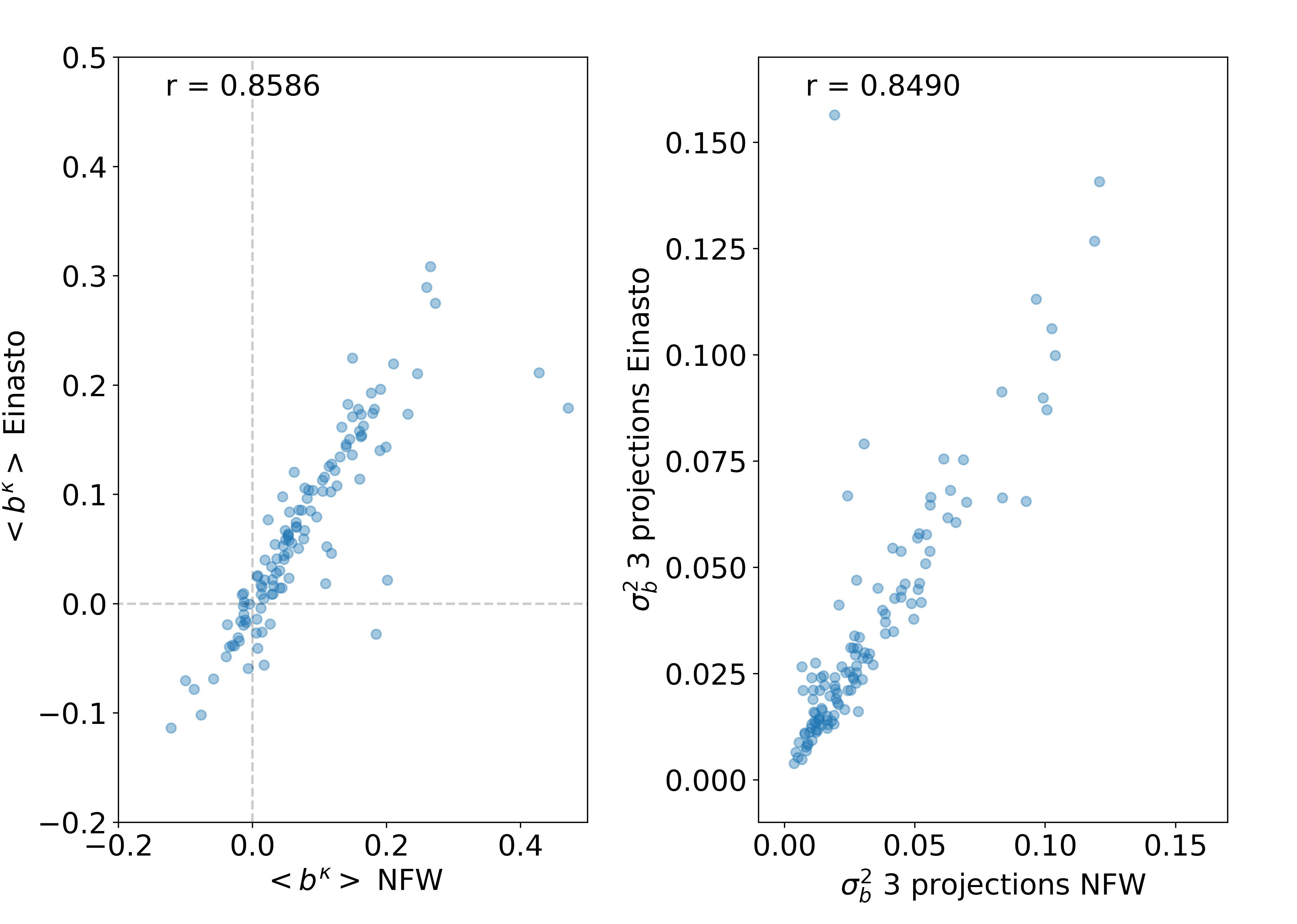}
        \caption{Mean bias of the three projections (left) and $\sigma_{b, \mathrm{3\; projections}}^2$ (right) per cluster for Einasto against NFW reconstructions. Each dot corresponds to one cluster. We give the Pearson correlation coefficient for the values given in each panel.}
        \label{fig:meanvsmean}
    \end{figure}    

    \begin{figure}
      \centering
      \includegraphics[trim={0pt 0pt 0pt 0pt}, scale=0.37]{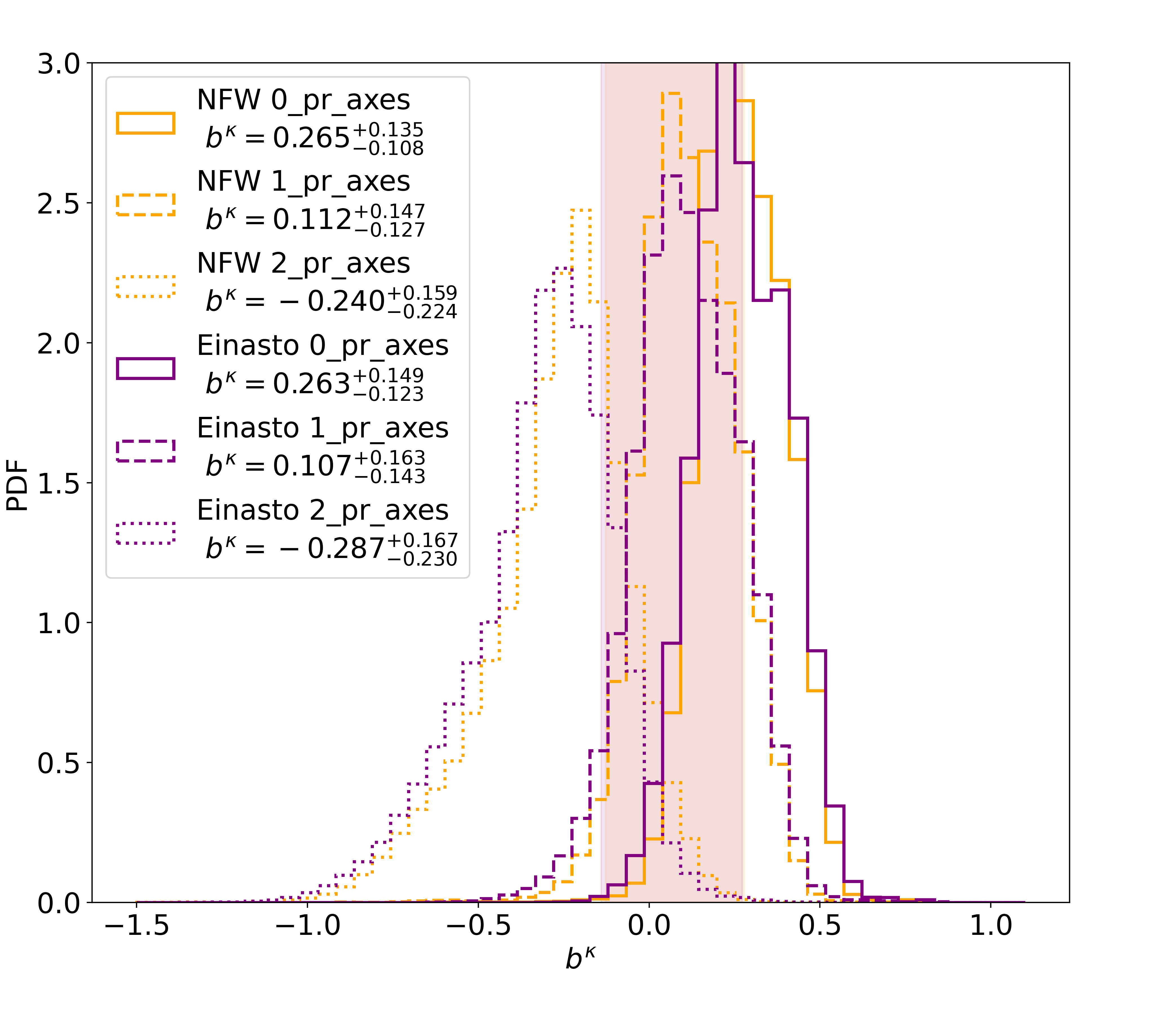}
      \caption{Mass bias for NFW (orange) and Einasto (purple) fits of the $\Sigma$-profiles when aligned with the principal axes of the clusters regarding their moments of inertia. Solid lines correspond to 0\_pr\_axes, dashed lines to 1\_pr\_axes, and dotted lines to 2\_pr\_axes. We give the value of the median and the 16th and 84th percentiles. Shaded areas represent the $1\sigma$ regions of the bias obtained for all the clusters with the three random projections (Fig.~\ref{fig:kappabias}).}
      \label{fig:histpraxes}
    \end{figure}  
    We present in Fig.~\ref{fig:kappabias} the bias estimated from all the analysed maps with respect to the true $M_{500}$ for each cluster. Orange and purple markers show the bias and uncertainties measured for the NFW and Einasto mass density models, respectively, with the uncertainties from the fits. 
    The horizontal orange line shows the mean bias for NFW, while the purple corresponds to Einasto.
    Einasto gives, on average, a less biased estimate of the mass of the clusters. The median biases for NFW and Einasto reconstructions are med$(b^{\kappa}) = 0.097$ and 0.086, respectively. We identify the origin of the bias from a thorough examination of the $\Sigma$-profiles presented in Sect.~\ref{sec:massreconstruction}.

    We present in Fig.~\ref{fig:nfweinastofits} the relative difference between the fit of the $\Sigma$-profile with NFW and Einasto models and the profiles measured from the convergence maps.
    We observe that for both models the mass density profiles are systematically underestimated for the radial ranges close to $R_{500}$. This is then translated into biased mass profiles and, consequently, biased $M_{500}$ estimates.

    One could think of improving the fits either by adapting the radial range ad hoc to get a less biased result or by choosing a more flexible density model. As aforementioned, this type of studies have already been done in the literature \citep{meneghetti2014, giocoli2012} and we do not go through those refinements in this work. In \citet{cui2018}, the authors concluded that the NFW model is a good representation of the spherical mass density profile of the clusters in \textsc{The Three Hundred} simulation, even considering different inner radii. But in \citet{giocoli2023}, the authors showed that \textsc{The Three Hundred} clusters do not follow perfect NFW density profiles.

    In the left panel in Fig.~\ref{fig:meanvsmean} we present the mean bias per cluster obtained from the NFW density fits with respect to the mean bias from the Einasto fits. There is a strong correlation between the NFW and Einasto biases, with a Pearson correlation coefficient of 0.86. This confirms that the impact of the model choice is not enough to blur the information in the convergence maps.

    We present in Fig.~\ref{fig:histpraxes} the bias measured from the maps projected along the three axes of inertia and verify that the orientation of the cluster affects directly the bias: when the cluster is elongated along the line of sight (2\_pr\_axes) the reconstructed mass is on average overestimated and when the major axes are on the plane of the sky (0\_pr\_axes), underestimated. In \cite{herbonnet2022}, the authors investigated the relation between the orientation of haloes and their brightest cluster galaxy (BCG). They found that the major axes of haloes and BCGs are aligned with an average separation of $\sim$ 20 degrees. For this reason, having access to the orientation of the BCG would provide a way to improve the knowledge of the orientation of the cluster, and therefore, of its mass reconstruction.
    
    Given all the mentioned differences between our work and the analysis in \citet{giocoli2023}, comparisons to the latter have to be done with extraordinary care. By fixing the truncation parameter of the tNFW model to $t=3$, they obtained that the weak lensing masses reconstructed for the clusters at redshift $z=0.22$ are on average biased low by $\sim 7\%$, with a standard deviation of 0.24 over the full sample (using the full \textsc{The Three Hundred} sample of clusters). Assuming also a fixed concentration parameter $c_{200}=3$ and considering the clusters at redshifts below $z=0.7$, according to \citet{giocoli2023}, weak lensing $M_{200}$ masses are on average underestimated by less than $5\%$. We observe in figure 14 in \citet{giocoli2023} that, for the redshift range considered in our work ($0.5 < z <0.9$), the mean biases on $M_{200}$ vary from $0\%$ to $20\%$ depending on the chosen value for the $c_{200}$ parameter. These results match our findings at $R_{500}$. Regarding the impact of the orientation of clusters on the mass bias, in \citet{giocoli2023} the authors concluded that when clusters are oriented along the major and minor axes, the mass biases are $\sim 25\%$ larger and smaller, respectively, with respect to the biases for random orientations. Therefore, their results are completely in line with Fig.~\ref{fig:histpraxes}.


\section{Scatter of the mass bias}
\label{sec:uncertainties}

In the previous section we have been looking at the mean biases estimated with each model for our full sample. The dispersion from cluster to cluster and for the different projections of the same cluster appear to be very important (Fig.~\ref{fig:kappabias}). In this section we try to identify and quantify the different effects that contribute to the scatter of the biases in the reconstruction. For this, we use an statistical approach and we consider the full sample.

We first define the total variance per model from the scatter of the bias values obtained with the same model for all the clusters and all their projections (see Fig.~\ref{fig:kappabias}):
\begin{equation}
  \sigma_{b, \mathrm{tot}}^2 = \frac{1}{N_{\mathrm{chains}}N_{\mathrm{projections}}N_{\mathrm{clusters}}}\sum_{i, j, k = 1, 1, 1}^{N_{\mathrm{chains}}, N_{\mathrm{projections}}, N_{\mathrm{clusters}}}(b_{i, j, k}^{\kappa} - <b^{\kappa}>)^{2},
  \label{eq:tot}
\end{equation}
where $N_{\mathrm{chains}}$ is the amount of chains kept after each fit, $N_{\mathrm{projections}}$ is the number of different projections per cluster and $N_{\mathrm{clusters}}$ the amount of clusters considered. Here $<b^{\kappa}>$ is the mean bias for all the clusters, accounting for all the projections and chains.

We separate this total variance in two main contributions: 1) the variance from cluster to cluster, $\sigma_{b, \mathrm{cluster-to-cluster}}^2$, showing how different are the bias values depending on the cluster, and 2) the variance of the results for each cluster, due to the different values depending on the projection, $\sigma_{b, \mathrm{3 \; projections}}^2$. So we write \citep[as in][]{bartalucci2023}:
\begin{equation}  
   \sigma_{b, \mathrm{tot}}^2 \sim \sigma_{b, \mathrm{cluster-to-cluster}}^2 + \sigma_{b, \mathrm{3 \; projections}}^2.
  \label{eq:projtot}
\end{equation}

Here $\sigma_{b, \mathrm{3 \; projections}}^2$ is the dispersion of the results of three projections, that for each cluster $c$ is obtained as,
\begin{equation}
  (\sigma_{b, \mathrm{3 \; projections}}^2)_c = \frac{1}{N_{\mathrm{chains}}N_{\mathrm{projections}}}\sum_{i, j = 1, 1}^{N_{\mathrm{chains}}, N_{\mathrm{projections}}}(b_{i, j, c}^{\kappa} - <b^{\kappa}>_{c})^{2},
  \label{eq:proj}
\end{equation}
where $ <b^{\kappa}>_{c}$ is the mean bias per cluster $c$, considering the different projections and all the chains.

At the same time, in each $\sigma_{b, \mathrm{3 \; projections}}^2$ per cluster there are two sources of dispersion that we name: 1) $\sigma_{b, \mathrm{intrinsic \; proj}}^2$, which accounts for the fact that the results obtained from three projections differ, and, 2) $\sigma_{b, \mathrm{fit}}^2$ that quantifies the uncertainties of the fits. Therefore, we can write
\begin{equation}
  \label{eq:projbias}
  \sigma_{b, \mathrm{3\; projections}}^2 \sim \sigma_{b, \mathrm{intrinsic \; proj}}^2 + \sigma_{b, \mathrm{fit}}^2 .
\end{equation}
In the following subsections we study each contribution to the total scatter.

\subsection{Fit uncertainties: $\sigma_{b, \mathrm{fit}}^2$}
The most evident source of scatter is the one related to the model fitting uncertainties, that is, the scatter propagated from the posterior distributions of the parameters in the fit of the convergence map at one specific projection of one cluster. We define $(\sigma_{b, \mathrm{fit}}^2)_{\mathrm{proj}, c}$ as the variance of the bias values of each map fit (`proj') for each cluster ($c$):
\begin{equation}
  \label{eq:fit}
  (\sigma_{b, \mathrm{fit}}^2)_{\mathrm{proj}, c} =  \frac{1}{N_{\mathrm{chains}}}\sum_{i= 1}^{N_{\mathrm{chains}}}(b_{i,\mathrm{proj},c}^{\kappa} - <b^{\kappa}>_{\mathrm{proj},c})^{2}.
\end{equation}  

We give the mean and median values in Table~\ref{tab:summarysigmas} (see Fig.~\ref{fig:scattersboth} for their distribution). On average 8\% ($\sigma_{b,  \mathrm{fit}}^2 \sim 0.007$) and 9\% ($\sigma_{b,  \mathrm{fit}}^2 \sim 0.008$) error in the mass reconstruction comes only from the uncertainty in the fit for the NFW and Einasto models, respectively.

\renewcommand{\arraystretch}{1.4}
\footnotesize
    \begin{table*}[]
      \centering
        \caption{Summary of the mass bias scatters studied in this work. We present the name given to each scatter term, the reference to the definition in the text and the obtained values for masses reconstructed using NFW and Einasto models.}
      \centering
      \footnotesize
        \begin{tabular}{c|c|c|c}
          \hline
          \hline
            Name & Definition &  NFW & Einasto \\ \hline

            $\sigma_{b, \mathrm{fit}}^2$ & Eq.~\ref{eq:fit} & $\left< \sigma_{b, \mathrm{fit}}^2 \right> =$ 0.0070 & $\left< \sigma_{b, \mathrm{fit}}^2 \right> =$ 0.0080 \\
             ~ & ~ &  med$(\sigma_{b, \mathrm{fit}}^2) =$ 0.0056 &  med$(\sigma_{b, \mathrm{fit}}^2) =$ 0.0055   \\\hline
            
            $\sigma_{b, \mathrm{3 \; projections}}^2$ & Eq.~\ref{eq:proj} & $\left< \sigma_{b, \mathrm{3 \; projections}}^2 \right> =$ 0.0326&  $\left< \sigma_{b, \mathrm{3 \; projections}}^2 \right> =$ 0.0356\\
            ~ & ~&  med$(\sigma_{b, \mathrm{3 \; projections}}^2) =$ 0.0251 &  med$(\sigma_{b, \mathrm{3 \; projections}}^2) =$ 0.0253 \\\hline  

            $\sigma_{b, \mathrm{intrinsic \; proj}}^2$ & Eq.~\ref{eq:projbias} &$  \left< \sigma_{b, \mathrm{3\; projections}}^2\right> - \left< \sigma_{b, \mathrm{fit}}^2\right> = $ 0.0256 & $  \left< \sigma_{b, \mathrm{3\; projections}}^2\right> - \left< \sigma_{b, \mathrm{fit}}^2\right> = $ 0.0276\\
            ~ & ~ &   med$(\sigma_{b, \mathrm{3\; projections}}^2) - $med$( \sigma_{b, \mathrm{fit}}^2) = $ 0.0195 &   med$(\sigma_{b, \mathrm{3\; projections}}^2) - $med$( \sigma_{b, \mathrm{fit}}^2) =$ 0.0198\\\hline
                       
            $\sigma_{b, \mathrm{tot}}^2$ & Eq.~\ref{eq:tot} & 0.0413 & 0.0428 \\\hline

            $\sigma_{b, \mathrm{cluster-to-cluster}}^2$ & Eq.~\ref{eq:projtot} & $ \sigma_{b, \mathrm{tot}}^2 - \left< \sigma_{b, \mathrm{3 \; projections}}^2\right>$ = 0.0087  & $ \sigma_{b, \mathrm{tot}}^2 - \left< \sigma_{b, \mathrm{3 \; projections}}^2\right>$ = 0.0072\\
            ~& ~ &  $ \sigma_{b, \mathrm{tot}}^2 - $med$( \sigma_{b, \mathrm{3 \; projections}}^2)$ = 0.0162 &  $ \sigma_{b, \mathrm{tot}}^2 - $med$( \sigma_{b, \mathrm{3 \; projections}}^2)$ = 0.0175 \\\hline
                      
            $\sigma_{b, \mathrm{mean}}^2$ & Eq.~\ref{eq:mean} & 0.0087 & 0.0071 \\\hline
     
            \hline
            \hline
             $(\sigma_{b, \mathrm{intrinsic \; proj}}^{\kappa\mathrm{-map}})^2$ & Eq.~\ref{eq:unavproj} &   \multicolumn{2}{|c}{From maps $\sim 0.01$}\\\hline
            
        \end{tabular}
        \vspace*{0.2cm}
    \begin{tablenotes}
    \centering
    \small
    \item \textbf{Notes.} For some cases we give the mean, $\left< ... \right>$, and median, med(...), values. 
    \end{tablenotes}
     
     \vspace*{0.2cm}
       \label{tab:summarysigmas}
        
    \end{table*}
     
\normalsize

\subsection{Scatter from 3 projections: $\sigma_{b, \mathrm{3\; projections}}^2$}
\label{sec:projection}
For each cluster three different projections are available and the variance of the results from the three is the $\sigma_{b, \mathrm{3 \; projections}}^2$ defined in Eq.~\ref{eq:proj}. As for the fit uncertainties, we give the mean and median values in Table~\ref{tab:summarysigmas} (the histograms of the variance for each cluster, accounting for the posterior distributions of the three projections together are shown in Fig.~\ref{fig:scattersboth}). In the right plot of Fig.~\ref{fig:meanvsmean} we present the $\sigma_{b, \mathrm{3 \; projections}}^2$ obtained with NFW and Einasto for each cluster in our sample, showing the correlation between the variances measured with the different models.

We study this scatter to quantify the uncertainty that should be added to observational mass reconstructions for which clusters are only observed in one projection. To truly estimate the projection effect, the dispersion of the masses obtained from the infinite projections of each cluster would be needed. In Appendix~\ref{sec:validation100} we instead compared the scatter obtained from 100 projections with the scatter obtained from the 3 simulation axes and from the 3 inertial axes and found that using the projections along $x, y$, and $z$ (i.e. 0, 1, and 2) returns a more similar scatter level to the 100 projections. The projections along the inertia axes instead exhibit a significant larger scatter. For this reason, in the scatter calculations we did not make use of the maps projected along the inertia axes.

\subsection{Intrinsic projection effect: $\sigma_{b, \mathrm{intrinsic \; proj}}^2$}
\label{sec:unavoidable}

For both NFW and Einasto models, there is a contribution to $\sigma_{b, \mathrm{3\; projections}}^2$ that is not explained by the uncertainties on the fits. We consider that this is due to the real `intrinsic projection effect' presented in Eq.~\ref{eq:projbias}. From the difference between the mean (and also between the median) values of $\sigma_{b, \mathrm{3\; projections}}^2$ and $\sigma_{b,\mathrm{fit}}^2$, the intrinsic projection effect is of the order of $\sim 0.020 - 0.025$. This corresponds to about $14- 16\%$ error on the mass bias and then on $M_{500}$. We stress that this is true for both NFW and Einasto, which is comforting, since we are looking for an intrinsic effect that should depend only on the clusters themselves.

\subsubsection*{Model independent intrinsic projection effect}

\begin{figure}
  \centering
  \includegraphics[scale=0.29]{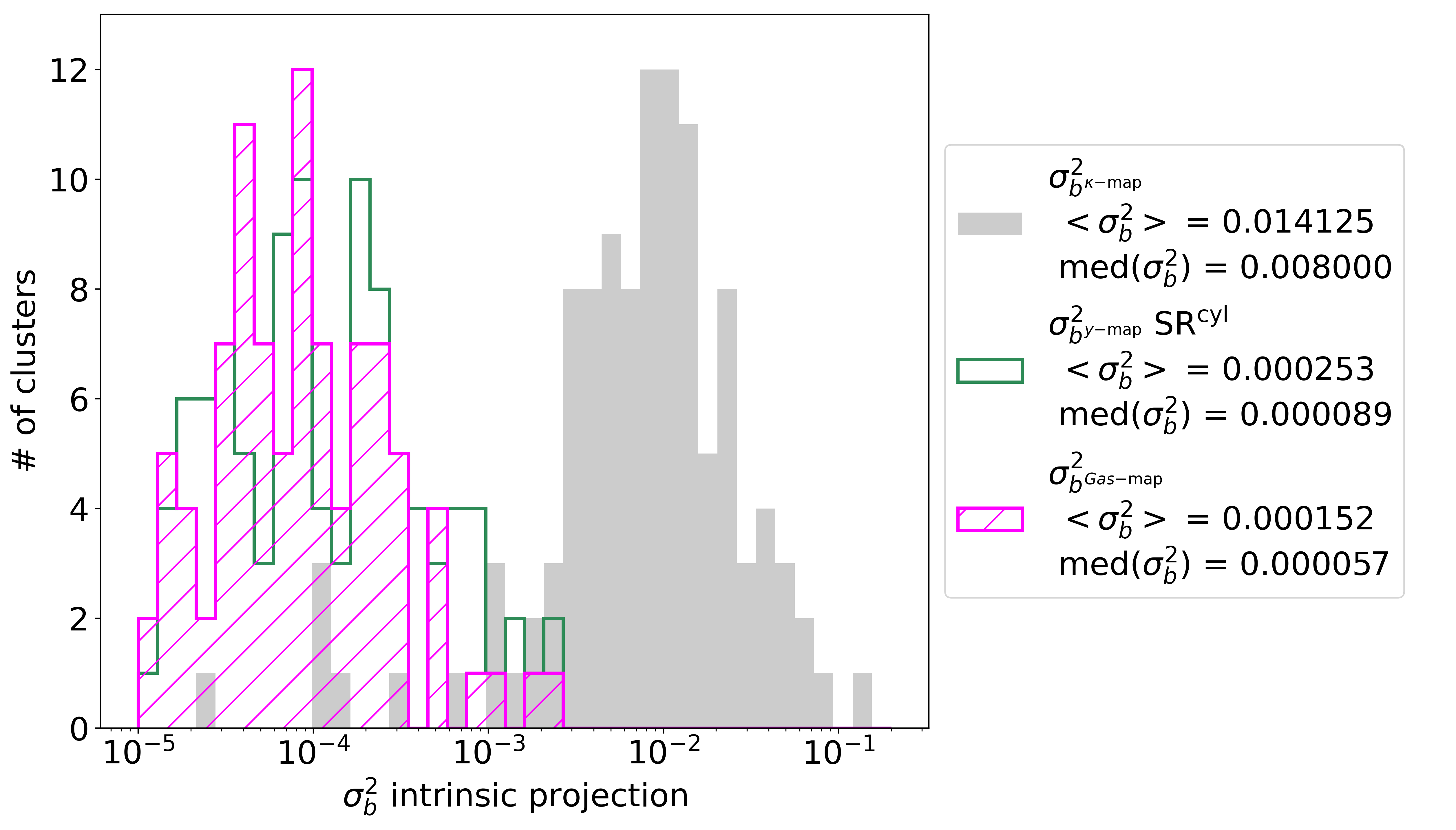}
  \caption{Variance of the mass bias estimated from maps. The grey histogram is obtained from the integration of convergence maps, i.e. $M_{500}^{\kappa, \mathrm{cyl}}$. The magenta histogram corresponds to the variance for gas masses ($M_{500}^{Gas\mathrm{, cyl}}$) estimated from gas mass maps. The empty green histogram shows the dispersion of the $M_{500}^{\mathrm{cyl}}$ after conversion from the $Y_{500}^{\mathrm{cyl}}$ measured from $y$-maps, using the scaling relation in Fig.~\ref{fig:SR}.}
  \label{fig:intrinsicproj}
\end{figure}

We are also interested in quantifying the contribution of $\sigma_{b, \mathrm{intrinsic \; proj}}^2$ without being affected by modelling effects. For this purpose, we take a different approach and infer masses directly from the projected convergence maps. By integrating the $\Sigma$-maps up to the true $\theta_{500}$, we compute directly the total mass of clusters from cylindrical integration, $M_{500}^{\kappa, \mathrm{cyl}}$, where $\tan(\theta_{500}) =  R_{500} /\mathcal{D}_{\mathrm{A}}$, with $R_{500} $ the true radius of the cluster and $\mathcal{D}_{\mathrm{A}}$ the angular diameter distance at the cluster redshift. For each cluster we have three different $M_{500}^{\kappa, \mathrm{cyl}}$, one per projected map. The grey histogram in Fig.~\ref{fig:intrinsicproj} shows the variance of the three mass biases per cluster $c$ for all the clusters in our sample obtained from
\begin{equation}
  (\sigma_{b, \mathrm{intrinsic \; proj}}^{\kappa\mathrm{-map}})^2_c = \frac{1}{N_{\mathrm{projections}}}\sum_{j = 1}^{N_{\mathrm{projections}}}(b_{j, c}^{\kappa\mathrm{-map}} - <b^{\kappa\mathrm{-map}}>_{c})^{2}.
  \label{eq:unavproj}
\end{equation}
On average, the projection effect when integrating the total mass at a fixed aperture is of the order of $(\sigma_{b, \mathrm{intrinsic \; proj}}^{\kappa\mathrm{-map}})^2 \sim 0.01$, therefore, an error of 10\% on the mass. The projection effect computed in this way is close, but not enough, to fully explain the $\sim$~15\% uncertainties obtained from the $ \sigma_{b, \mathrm{3\; projections}}^2 - \sigma_{b, \mathrm{fit}}^2$ difference.

An additional term could explain the extra dispersion: while the masses used to compute $\sigma_{b, \mathrm{3 \; projections}}^2$ are estimated at different radii (each $M_{500}^{\kappa}$ measured at the corresponding $R_{500}^{\kappa}$), here all the  $M_{500}^{\kappa, \mathrm{cyl}}$ for $(\sigma_{b, \mathrm{intrinsic \; proj}}^{\kappa\mathrm{-map}})^2$ are measured at fixed $\theta_{500}$ per cluster. Therefore $(\sigma_{b, \mathrm{intrinsic \; proj}}^{\kappa\mathrm{-map}})^2$ needs a corrective term to be fully comparable to $\sigma_{b, \mathrm{intrinsic \; proj}}^2$. To quantify this effect, we produced mock mass density profiles of galaxy clusters with a large variety of shapes and masses. We compared the scatter when estimating the $M_{500}$ spherically and up to the $R_{500}$ corresponding to each profile and the scatter when measuring the mass cylindrically integrated at a fixed aperture. All the details are given in Appendix~\ref{sec:cyl_vs_sph}. In conclusion, this additional variance can be estimated of the order of $\sim 0.001$ to $0.003$, so $\sim 3 - 5\%$, depending on the shape, redshift, and mass of the density profile of the cluster. We note that additional differences could appear when integrating spherically against cylindrically, related, for example, to the presence of structures. 

Therefore, both from a model dependent and a model independent approach we obtain compatible values for $\sigma_{b, \mathrm{intrinsic \; proj}}^2$. We conclude that when deriving cluster masses from observables tracing the total mass (for instance, the lensing effect), the intrinsic effect of the projection creates an error in the $M_{500}$ estimates between 10\% and 16\%. These values could vary with the depth of the considered $\kappa$-maps. 

\subsection{Total scatter: $\sigma_{b, \mathrm{tot}}^2$}

Finally, we quantify the total variance of the bias along the sample and the scatter from cluster to cluster. We compute, following the Eq.~\ref{eq:tot}, the total variance of the bias in our sample by accounting for all the clusters, with all the projections and their fit uncertainties. Following Eq.~\ref{eq:projtot}, we obtain the excess with respect to the projection and fit effects. All the values are given in Table~\ref{tab:summarysigmas}.

The total dispersion appears to be larger than the average projection effect. This could be due to some clusters with a very important projection effect (right panel in Fig.~\ref{fig:meanvsmean}) that might be affecting significantly $\sigma_{b, \mathrm{tot}}^2$. But it could also originate from the scatter of the mean bias values per cluster. We compute the mean bias per cluster $c$ accounting for the masses reconstructed from all the chains of the three convergence maps ($<b^{\kappa}>_c = <( M_{500} - M^{\kappa}_{500} )/ M_{500} > $) and quantify how different are the mean biases for all the clusters in the sample. We define the variance of the mean biases, $\sigma_{b, \mathrm{mean}}^2$, as:
\begin{equation}
  \sigma_{b, \mathrm{mean}}^2 = \frac{1}{N_{\mathrm{clusters}}}\sum_{k = 1}^{N_{\mathrm{clusters}}}(<b^{\kappa}>_k - \left< <b^{\kappa}>_{c} \right>)^{2},
  \label{eq:mean}
\end{equation}
where $\left< <b^{\kappa}>_{c} \right>$ is the mean over the mean biases of all the clusters. Results are shown in Table~\ref{tab:summarysigmas}. The missing dispersion in $\sigma_{b, \mathrm{tot}}^2$ seems to be explained by the different mean bias values of different clusters, $\sigma_{b, \mathrm{cluster-to-cluster}}^2 \sim \sigma_{b, \mathrm{mean}}^2$. We notice that it is a model dependent effect, NFW results being slightly more scattered with respect to $\left< <b^{\kappa}>_{c} \right>$. This difference could mean that NFW reconstructions fail more often in the description of the mass of clusters and, compared to Einasto, NFW fit uncertainties are not large enough to account for it.

\begin{figure*}[h]
    \centering
    \includegraphics[scale=0.4]{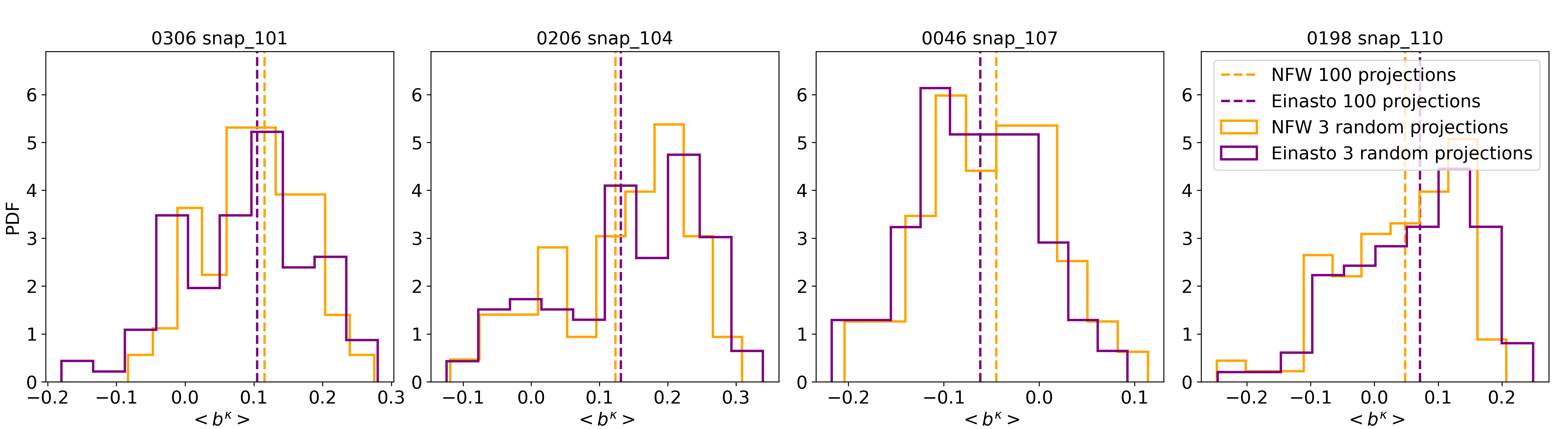}
    \caption{Mean bias per cluster. The vertical lines show the mean bias when using 100 projections to reconstruct the mass. The histograms are obtained from the mean bias computed by taking randomly 3 projections out of the 100 available 100 times.}
    \label{fig:100vs3}
\end{figure*}
\begin{figure*}[h]
       \centering
        \begin{minipage}[b]{0.48\textwidth}
        \includegraphics[scale=0.38]{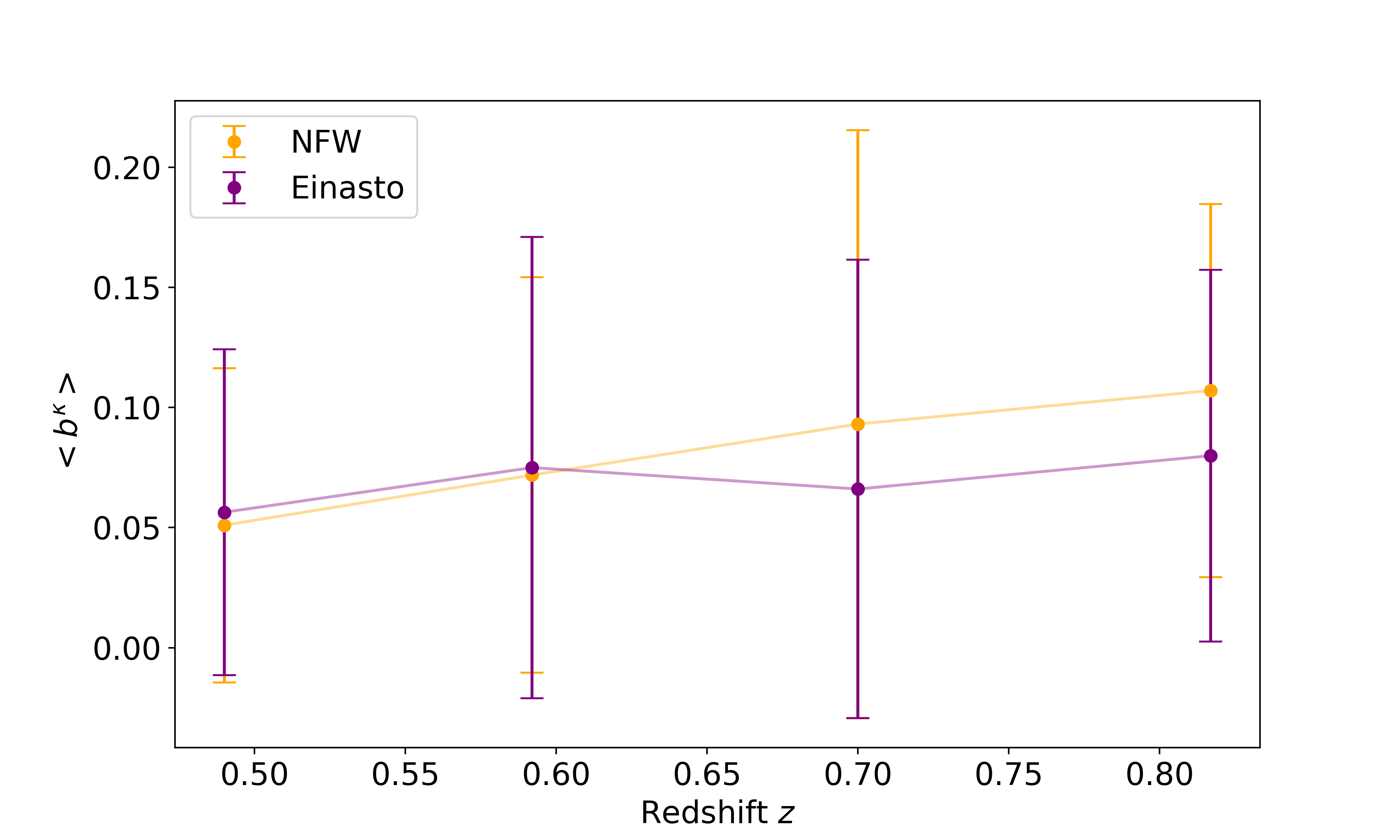}
        \end{minipage}
        \hfill
        \begin{minipage}[b]{0.50\textwidth}
        \includegraphics[scale=0.38]{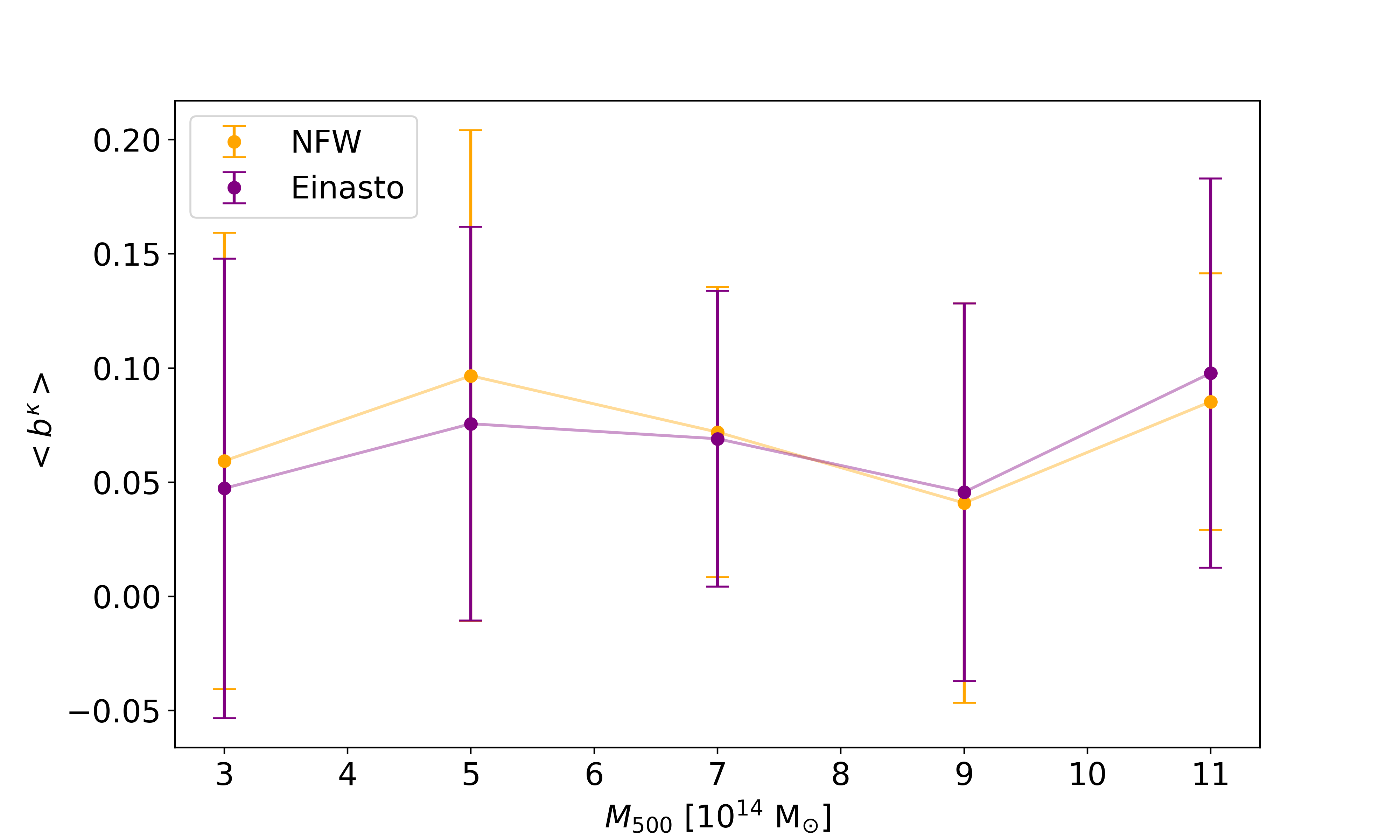}
        \end{minipage}
        \caption{Bias for clusters in different redshift (left) and mass (right) bins. We give the mean value and the $\sigma_{b, \mathrm{mean}}$ scatter for the mean bias of clusters in each bin. Results for NFW and Einasto models are shown in orange and purple, respectively, with a line as a guide to the eye.}
\label{fig:lpszbins}
\end{figure*}

\subsection{Intrinsic scatter: $\sigma_{b, \mathrm{cluster-to-cluster}}^2$}
The dispersion of the bias from cluster to cluster can be caused by multiple effects. We investigate here the impact of data limitations and of the mass, redshift, and dynamical state of the clusters.

\subsubsection*{Data limitations}

We only consider 3 projections per cluster and the mean bias obtained from these 3 may not be a good representation of the true mean bias.
We test the impact of this effect using four clusters for which we have 100 projections (see Appendix~\ref{sec:validation100}). For each cluster we calculate the mean bias accounting for the 100 projections. Values are shown as vertical dashed lines in Fig.~\ref{fig:100vs3}. At the same time, we take randomly three projections out of the 100 and compute the mean for those trios. The histograms in Fig.~\ref{fig:100vs3} show the results for 100 realisations. They illustrate that the mean bias $<b^{\kappa}>_c$ estimated from 3 random projections can in some cases be significantly different from the true one. Although histograms are centred in the mean bias of 100 projections, their variance is of the order of $\sim$ 0.005 - 0.010, which also contribute to $\sigma_{b, \mathrm{cluster-to-cluster}}^2$ or $\sigma_{b, \mathrm{mean}}^2$.

\subsubsection*{Impact of mass and redshift}

There could be also a correlation between intrinsic properties of clusters (mass, redshift, etc.) and the measured biases, meaning different responses to the model fitting depending on cluster properties. To investigate these effects, we separate our sample in subsamples. We choose a binning close to the one used for the NIKA2 LPSZ (Sect.~\ref{sec:lpsz}): four bins in redshift (one per snapshot) and five bins in mass (in the NIKA2 LPSZ only two bins in redshift are considered, but with the objective of observing any evolution, we choose to distinguish four). The five bins in mass are:  $M_{500}/ 10^{14} \mathrm{M}_{\odot} < 4$,  $ 4 \leq M_{500}/  10^{14} \mathrm{M}_{\odot} < 6 $,  $ 6  \leq M_{500}/ 10^{14} \mathrm{M}_{\odot} < 8$,  $ 8 \leq M_{500}/ 10^{14} \mathrm{M}_{\odot} < 10$, and  $ 10 \leq M_{500}/ 10^{14} \mathrm{M}_{\odot}$.
        
We show in Fig.~\ref{fig:lpszbins} the evolution of the mean bias with redshift and mass for NFW and Einasto mass reconstructions. A slight evolution of the bias with redshift is observed for NFW results, but not for Einasto. We note, in any case, that our results are compatible with no evolution. In \citet{giocoli2023}, the authors observe an evolution of the bias of $M_{200}$ masses with redshift and claim that such evolution is due to the low weak lensing signal-to-noise at high redshift. In the right panel in Fig.~\ref{fig:lpszbins}, we observe no evolution of the bias with the true $M_{500}$ either.

\subsubsection*{Impact of the dynamical state}
\label{sec:dynstate}
\begin{figure}
        \begin{minipage}[b]{1\textwidth}
        \includegraphics[trim = {0pt 0pt 0pt 0pt}, scale=0.34]{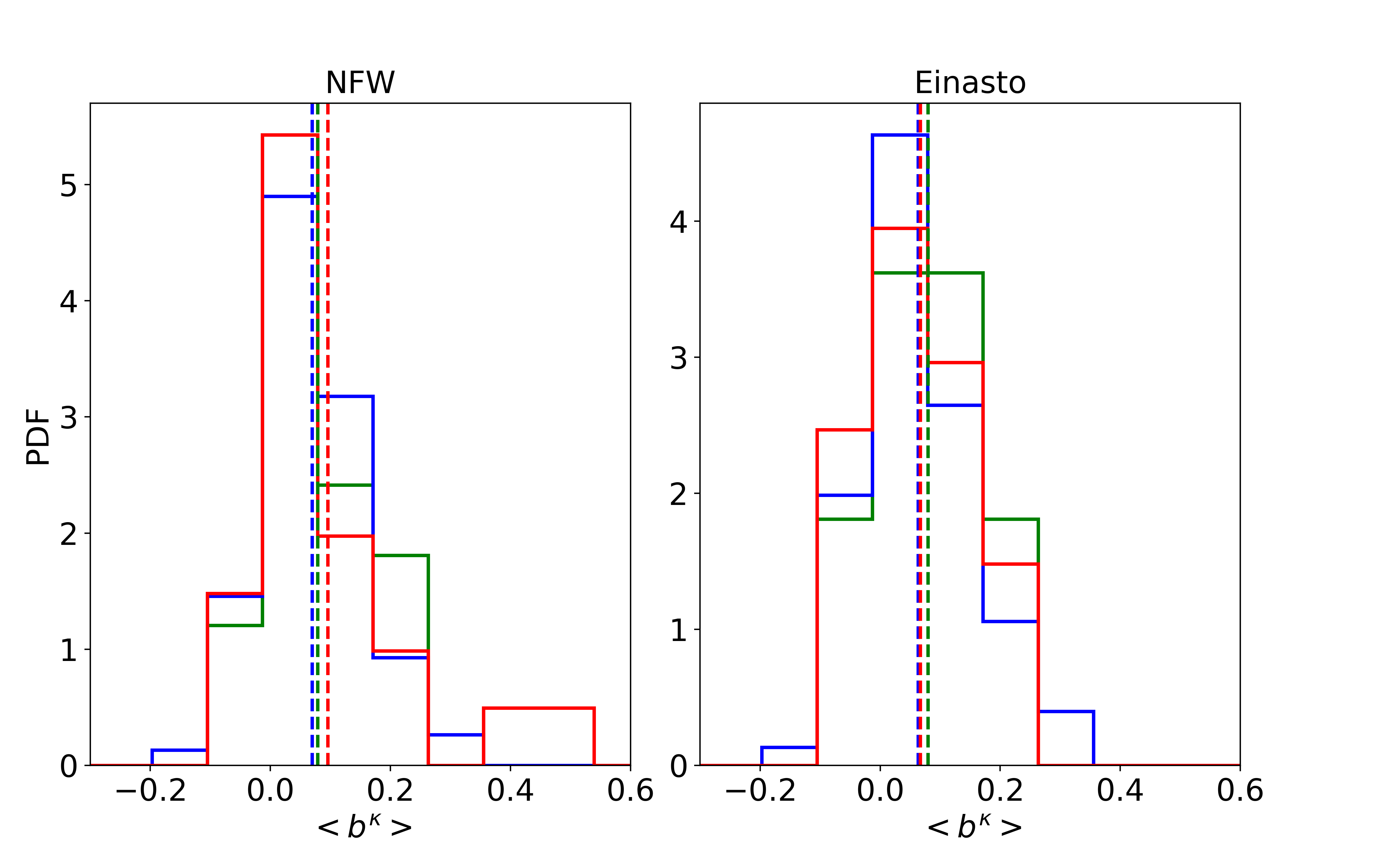}
        \end{minipage}
        \hfill
        \begin{minipage}[b]{1\textwidth}
        \includegraphics[trim = {0pt 0pt 0pt 0pt}, scale=0.34]{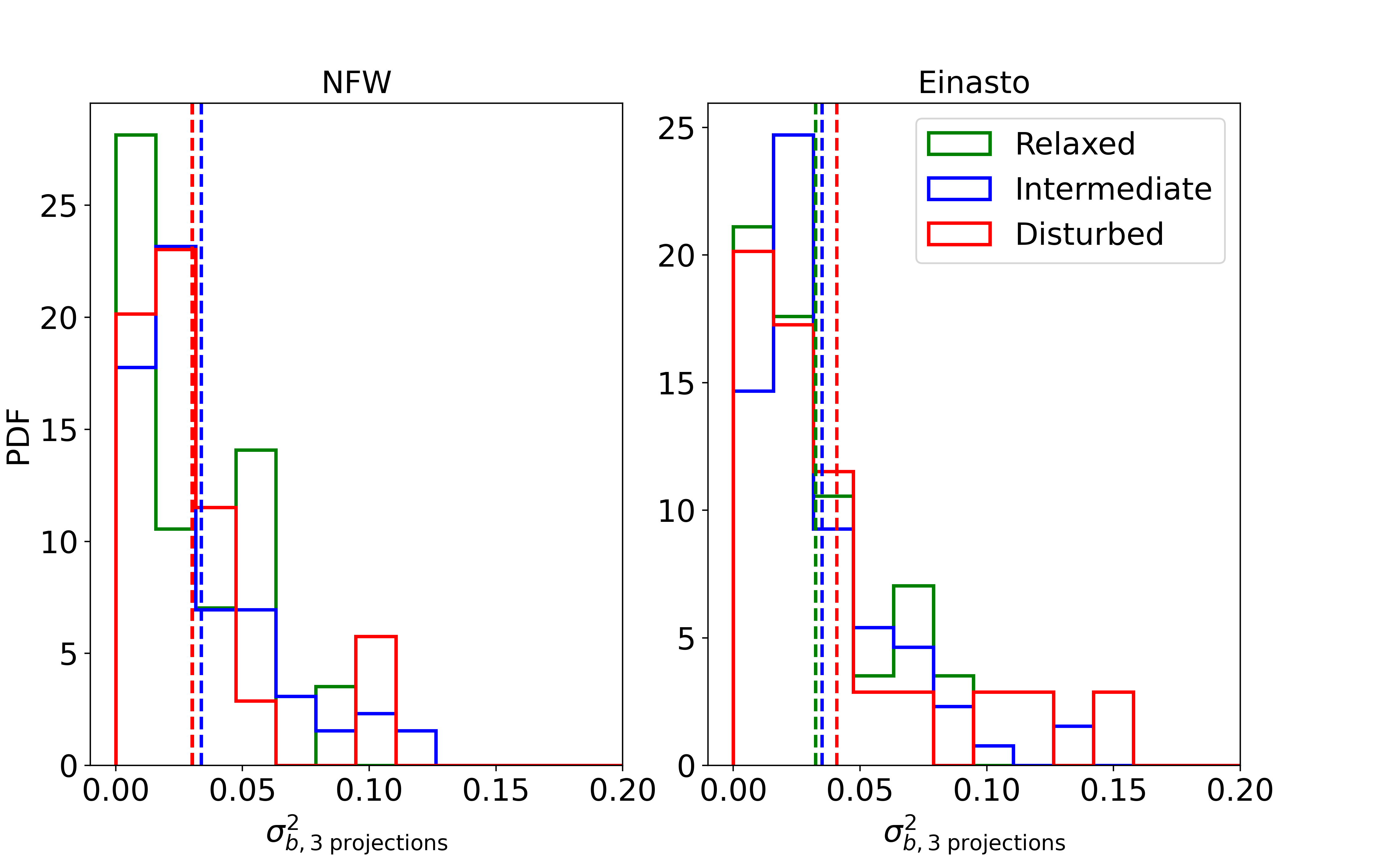}
        \end{minipage}
        \caption{Bias properties depending on the dynamical state of clusters within $R_{200}$. Top: distributions of the mean bias per cluster for relaxed (green), intermediate (blue), and disturbed (red) clusters. We show the bias from NFW (left) and Einasto (right) reconstructions. Bottom: distributions of the variance of the bias per cluster accounting for three projections. Clusters are classified according to their dynamical state and results for NFW (left) and Einasto (right) are shown. The vertical lines show the mean values of the distributions.}
\label{fig:morpho}
\end{figure}  
  \begin{figure}
        \centering
        \includegraphics[scale=0.33]{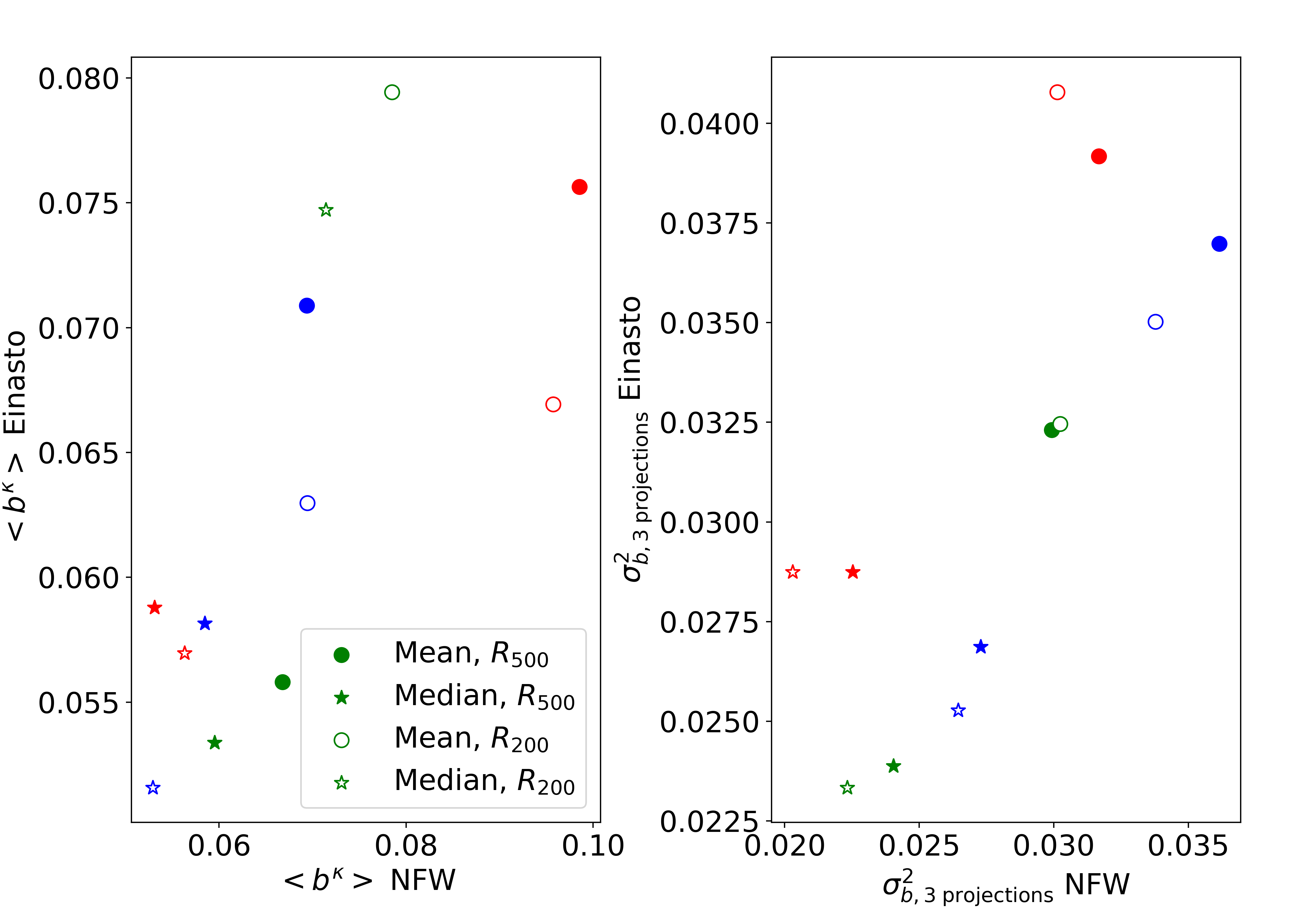}
        \caption{Summarised bias properties depending on the dynamical state of clusters within $R_{200}$ (empty) and $R_{500}$ (filled). Left: mean (circle) and median (star) of the mean biases of each type of cluster. Right: mean and median of the $\sigma^{2}_{b,\; 3\; \mathrm{projections}}$ for all the clusters of each type. We show in green, blue, and red the results for relaxed, intermediate, and disturbed clusters, respectively.}
        \label{fig:allmorpho}
    \end{figure}
The dynamical state of clusters can also be an indicator of the bias of masses reconstructed from projected maps and their dispersion. To classify the clusters in our sample according to their dynamical state, we use the $f_s$ and $\Delta_{r}$ indicators at $R_{200}$ (and $R_{500}$) that correspond respectively to the ratio between the sub-halo masses and the cluster mass within $R_{200}$ ($R_{500}$) aperture and the offset between the centre of mass of the cluster and the maximum density peak position, normalised by $R_{200}$ ($R_{500}$). Following the definitions in \citet{deluca2021}, clusters are relaxed (disturbed) if $f_s < 0.1$ and $\Delta_{r} < 0.1$ ($f_s > 0.1$ and $\Delta_{r} > 0.1$). Otherwise, we consider them in an intermediate dynamical state. 

In the top panels in Fig.~\ref{fig:morpho} we present the distribution of the mean bias per cluster independently for relaxed (green), intermediate (blue), and disturbed (red) clusters, classified with the indicators at $R_{200}$. The variance of these mean biases from cluster to cluster is larger for the disturbed sample ($\sigma^2_{b,\; \mathrm{mean}}=0.0176$ and $0.0070$ for NFW and Einasto, respectively) than for the relaxed one ($\sigma^2_{b,\; \mathrm{mean}}=0.0048$ for NFW and $0.0055$ for Einasto), but, on average, relaxed clusters are not less biased than disturbed ones.

In the bottom panels in Fig.~\ref{fig:morpho} we show the variance of the bias per cluster as computed from the three projections. Although we would expect to detect the departure from sphericity in the measurement of $\sigma_{b, \mathrm{3\; projections}}^2$, from these results we only observe such effect for the Einasto model. The three subsamples give scatter values that are compatible with those of the full sample. 

The classification of clusters changes if we use the indicators at $R_{500}$. In Fig.~\ref{fig:allmorpho} we summarise the mean and median values obtained from the distributions in Fig.~\ref{fig:morpho} for dynamical states of clusters defined within $R_{500}$ and $R_{200}$. From this figure, the bias of masses reconstructed with our method seems vaguely related to the relaxation status of the galaxy clusters. 

In \citet{giocoli2023}, the authors used a stronger constraint to classify clusters as relaxed: they considered relaxed systems if $f_s < 0.1$ and $\Delta_{r} < 0.05$ \citep[$x_{\mathrm{off}}$ in][]{giocoli2023} or if $f_s < 0.1$, $\Delta_{r} < 0.04$, and $0.85 < \eta < 1.15$ \citep[with $\eta$ the virial ratio, as in][]{cui2018}, all within $R_{200}$, and unrelaxed otherwise. Following this classification, they concluded that unrelaxed clusters are slightly worse modelled than relaxed ones, but that the mean bias of both types of clusters is completely consistent with that of the full sample.

\subsection{Summary}

In this section we have quantified the different contributions to the dispersion of mass estimates of clusters reconstructed from convergence maps. We find that $M_{500}$ cluster masses reconstructed from convergence maps have an uncertainty of the order of 9\% that corresponds to the uncertainty in the fit and varies with the chosen model. In addition, a 10 to 16\% uncertainty should be added to account for intrinsic projection effects. Regarding the variation of the bias along the cluster sample, which is of the order of 9\% ($\sigma_{b, \mathrm{cluster-to-cluster}}^2 = 0.009$ for NFW and $0.007$ for Einasto), we conclude that it is probably overestimated due to the limited projections per cluster available. When we try to correlate the bias and its dispersion to intrinsic characteristics of clusters, we find that redshift, mass, and dynamical state appear vaguely correlated to the bias. Additional effects that we have not considered in this work, such as the presence of substructures along the line of sight, could also be at the origin of the cluster-to-cluster scatter in observations. 

The average bias of the mass estimates reconstructed with the method presented in this paper, as well as the fit uncertainty, are directly impacted by the radial range that is considered in the fit. Depending on the size of the error bars of the $\Sigma$-profile at each radius and the ability of the density model to describe the mass at the desired radii, mass estimates can be more or less biased. On the contrary, the intrinsic projection effect (quantified by $\sigma_{b, \mathrm{intrinsic \; proj}}^2$, as well as by the comparison of the biases obtained for the 0\_pr\_axes, 1\_pr\_axes, and 2\_pr\_axes projections) remains of the same order, even for the extreme case where we consider all the radial ranges available for the fit of the mass density model.

\section{Comparison to gas observables}
\label{sec:gas}
As investigated in several works \citep[e.g.][]{meneghetti2010, rasia2012}, the spatial distribution of the gas in the ICM does not follow necessarily the same distribution as the rest of the matter. This implies that the aforementioned conclusions may not apply to masses reconstructed from gas observables. In this section, we compare the projection effect obtained for total matter observables to the effect for gas observables.

To measure the impact of the projection when estimating masses from gas maps we take two approaches. The most direct option consists in measuring the mass of the gas from the gas mass maps (Sect.~\ref{sec:gasmassmaps}). We compute $M_{500}^{Gas\mathrm{, cyl}}$ by integrating the maps up to $\theta_{500}$. The distribution of dispersions for the three masses calculated from the different projections is shown for all the clusters in the magenta histogram in Fig.~\ref{fig:intrinsicproj}. These results reflect the impact of the orientation as regards the arrangement of the gas in the cluster, but we note that $M^{Gas}$ is not an estimate of its total mass.

    A less direct option consists in estimating the mass from the integrated $Y_{500}$ signal in the $y\text{-maps}$ by applying a $Y_{500}-M_{500}$ scaling relation. First, the signal in the $y\text{-maps}$ is integrated up to the real $\theta_{500}$ for each cluster, which gives the cylindrically integrated Compton-$y$ parameter $Y_{500}^{\mathrm{cyl}}$. We then convert the $Y_{500}^{\mathrm{cyl}}$ in $M_{500}$ assuming a given scaling relation. 

    There are several scaling relations in the literature that relate $Y_{500}$ to $M_{500}$ \citep[A10,][]{cui2018}. However, most of them look for a relation between the observed $Y_{500}^{\mathrm{cyl}}$ and the spherical mass $M_{500}^{\mathrm{sph}}$. In our case, we build a $Y_{500}^{\mathrm{cyl}} - M_{500}^{\kappa,\mathrm{cyl}}  $ scaling relation for our sample that allows us to compare cylindrically integrated masses. This scaling relation, together with $Y_{500}^{\mathrm{cyl}} - M_{500}^{\mathrm{sph}}$ are shown in Fig.~\ref{fig:SR} in Appendix~\ref{sec:appendixD}. The dispersion of the recovered masses for different projections of a cluster allows us to quantify the intrinsic projection effect. The empty green histogram in Fig.~\ref{fig:intrinsicproj} shows the distribution of the dispersion of the biases estimated for all clusters from $y\text{-maps}$.
    
    We observe that both approaches suggest a $\sigma_{b, \mathrm{intrinsic \; proj}}^2 \sim 10^{-4}$ (a few percent scatter on $M_{500}$) for the gas, meaning that for the considered data and within $\theta_{500}$, the gas is an order of magnitude less disperse (i.e. more spherically distributed) than the dark matter also within the same $\theta_{500}$.  This is in agreement with previous works \citep{Buote_2012,Becker_2011, pratt2019,meneghetti2010, rasia2012} that also indicate that mass reconstructions from gas are less scattered than those from dark matter observables due to the difference in the spatial distribution of matter.

    From the comparison between the bias of cylindrically integrated masses from the $\kappa$-maps and the bias of masses obtained from $y$-maps (with the $Y_{500}^{\mathrm{cyl}} - M_{500}^{\kappa,\mathrm{cyl}}  $ scaling relation), we obtain that their correlation is weak, with a Pearson correlation coefficient of 0.26. The correlation is stronger between the bias dispersions (i.e. $\sigma_{b}$) measured from the three random projections in $\kappa$- and $y$-maps: r = 0.48.

\section{Summary and conclusions}
\label{sec:conclusions}
We have studied the bias of galaxy cluster mass estimates reconstructed by fitting three-dimensional NFW and Einasto density models to projected mass density profiles obtained from convergence maps. We have performed the analysis making use of \textsc{The Three Hundred} \texttt{GADGET-X} hydrodynamical simulation clusters, selected to be representative of the NIKA2 LPSZ sample. All the results shown here were obtained with the 122 clusters of the three twin samples combined (Sect.~\ref{sec:twinsamples}). We checked that conclusions do not vary from twin sample to twin sample. 

We decided to perform the fits of the mass density profiles accounting only for the radial ranges with uncertainties comparable to the ones of the profiles computed from CLASH convergence maps. Although the flexibility of the Einasto model permits fitting the projected density profiles by accounting for the different slopes, the fit fails describing the mass at $\sim R_{500}$, almost in the same way as the NFW model. Some works in the literature \citep{meneghetti2014} choose ad hoc the radial range and conclude that, as expected, models with more parameters fit better the density profile and therefore, give lower biases. The latter is also true for our case, but the difference between Einasto and NFW biases is not significant. Our main conclusion is that a projected mass density fit that is overall good does not give necessarily a good $M_{500}$ estimate.

Despite the slight differences between NFW and Einasto mass reconstructions, we have observed that both the mean bias and the scatter of the recovered masses are strongly correlated for the two models. This could mean that, in spite of the impact of the model choice, the effect is not enough to blur the information in the convergence maps. Quantitatively, the NFW and Einasto $M_{500}$ masses are biased by 8\% and 7\%, respectively. Therefore, both are in the 5$-$10 \% bias range obtained in previous simulation-based works \citep{rasia2012, meneghetti2014,Becker_2011,giocoli2023}.

Regarding the errors associated with the mass estimates, we consider different contributions. With our approach, the uncertainty of the density profile fit to the convergence map introduces a mean error on the mass of 8 and 9\% for NFW and Einasto profiles, respectively. An additional contribution comes from projection effects and we estimate that about 10 to 16\% of dispersion should be considered. Such result matches the conclusions for lensing masses in the \cite{pratt2019} review. Accounting for the projection effect together with the uncertainties of the fitting, the scatter is of the order of $\sim 18\%$, slightly less than the values in \cite{Becker_2011} and \cite{rasia2012}, the latter being obtained from the fit of tangential shear profiles of simulated clusters.

When accounting for the full sample of \textsc{The Three Hundred}-NIKA2 LPSZ clusters, there is an excess of dispersion with respect to projection effects. We investigate the origin of such dispersion and we find, firstly, an uncertainty due to the fact that we use only three projections per cluster. Making use of 100 random projections for four clusters, we verify that considering only 3 projections per cluster, the mean bias of the three can be scattered about up to 7-10\%.

We tried to correlate the bias to intrinsic properties of clusters and we checked the evolution of the mass bias with the true mass and redshift of clusters. Nevertheless, we did not find any evolutionary trend of the bias with the mass nor the redshift. In addition, we investigated the relation between the dynamical state and reconstructed masses of clusters, concluding that disturbed clusters are not particularly more biased than relaxed ones.

Regarding the orientation of clusters, we confirm that with our method and for clusters observed elongated along the line of sight, the reconstructed masses are overestimated. On the contrary, masses are underestimated if the major axes of clusters are on the plane of the sky. These conclusions are in agreement with the results in \citet{giocoli2023}, where \textsc{The Three Hundred} convergence maps are used to create weak lensing observables for the preparation of the \textit{Euclid} mission.

Given the similarities and differences between both works, it was indispensable to compare, all along this paper, our outcomes to \citet{giocoli2023}. Despite all the presented nuances, both works seem to converge towards resembling conclusions.

Finally, we also compared how spherical is the spatial distribution of total matter and gas by measuring the dispersion from projection to projection in total mass and gas maps. Within $R_{500}$ the gas is more spherically distributed than the dark matter, which allows one to have a mass reconstruction from projected maps less dependent on the orientation of the cluster. This was already known from previous studies \citep{meneghetti2010}, which motivated the observations of clusters in X-rays and SZ. For observational HSE-to-lensing mass biases estimated as for the LPSZ clusters (Sect.~\ref{sec:lpsz}), as well as for other works in the literature \citep[e.g.][]{bartalucci2018,serenoettoricomalit,Foex2012}, this would imply that most of the projection scatter is introduced by the lensing mass estimate.


\begin{acknowledgements}
 
This work has been made possible by \textsc{The Three Hundred} collaboration$^{\ref{the300ref}}$. MDP, AF, and AP acknowledge support from Sapienza Universit\a' di Roma thanks to Progetti di Ricerca Medi 2021, RM12117A51D5269B. WC is supported by the STFC AGP Grant ST/V000594/1 and the Atracci\'{o}n de Talento Contract no. 2020-T1/TIC-19882 granted by the Comunidad de Madrid in Spain. WC and GY thank the Ministerio de Ciencia e Innovación (Spain) for financial support under Project grant PID2021-122603NB-C21. This project was carried out using the Python libraries \texttt{matplotlib} \citep{Hunter2007}, \texttt{numpy} \citep{Harris2020}, \texttt{astropy} \citep{Astropy2013, Astropy2018}, and \texttt{emcee} \citep{foreman,goodman}.
\end{acknowledgements}


%
\bibliographystyle{aa} 
\bibliography{mybibliography} 
%

\begin{appendix}

\section{Mass reconstruction}
\label{sec:appendixA}

\subsection{Mass density model best-fit parameters}
\begin{figure*}[h]
        \centering
        \begin{minipage}[t]{0.4\textwidth}
        \includegraphics[trim={0pt 0pt 0pt 0pt},scale=0.33]{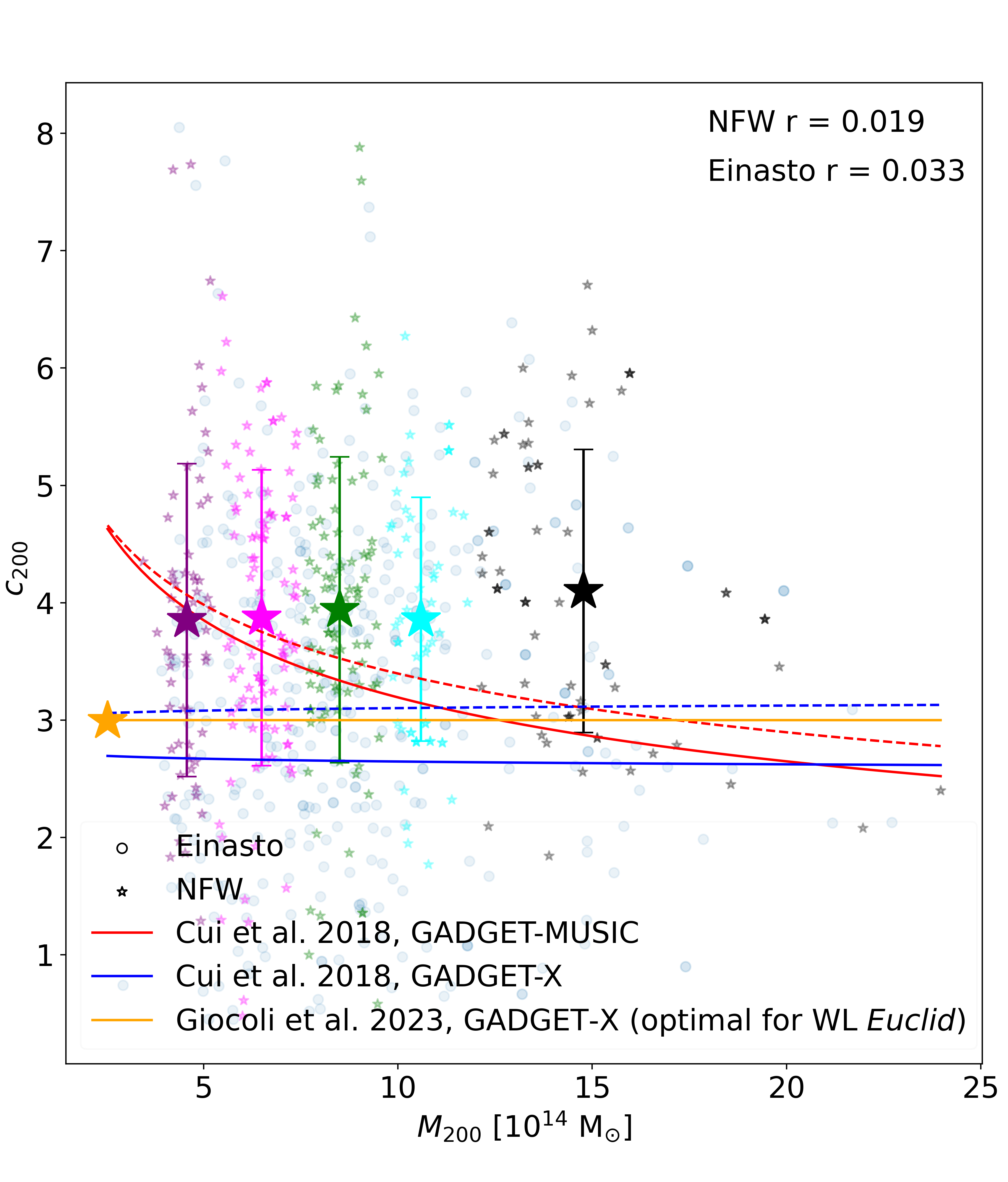}
        \end{minipage}
        \hfill
        \begin{minipage}[b]{0.59\textwidth}
        \includegraphics[trim={0pt 0pt 0pt 0pt},scale=0.33]{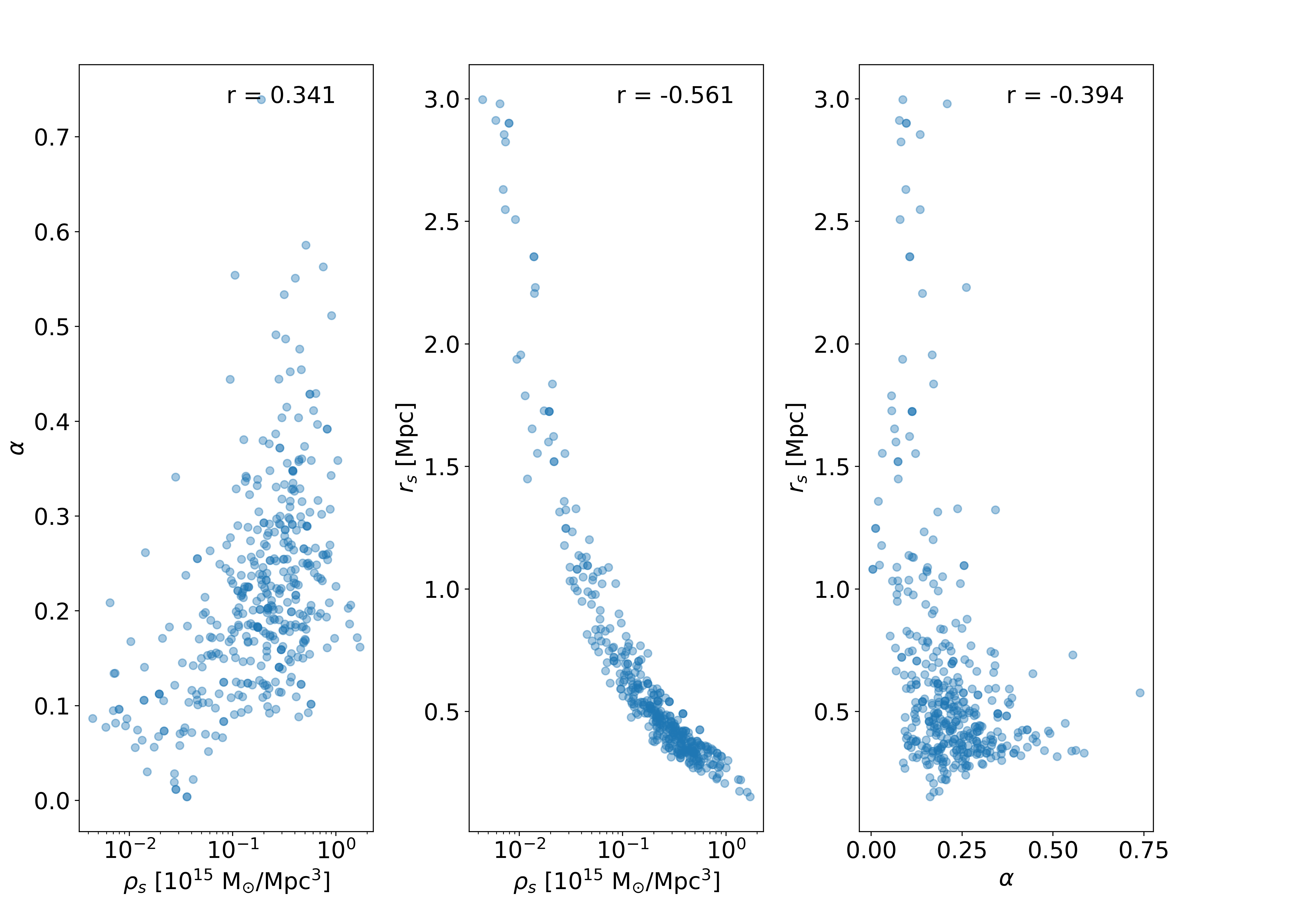}
        \end{minipage}
    \caption{Best-fit values of the posterior distributions of NFW (left) and Einasto (right) parameters. Each point corresponds to one projection of one cluster. We give the Pearson correlation coefficients for the parameters presented in each panel. In the left panel the big stars show the mean concentration and the standard deviation in different mass bins. Blue and red lines show the concentration-mass relations for \texttt{GADGET-X} and \texttt{GADGET-MUSIC} simulations obtained in \citet{cui2018} accounting for different radial ranges: with solid lines considering data above $0.05\times R_{200}$ and with dashed lines the results considering data above $34 h^{-1}$~kpc. The horizontal orange line at $c_{200} = 3$ in the left panel represents the best concentration value obtained in \citet{giocoli2023} to reconstruct unbiased masses. For comparison, in the left panel we also plot the $M_{200}$ and $c_{200} = R_{200}/r_{s}$ obtained for the Einasto best-fits. }
    \label{fig:fitparams}
\end{figure*}

In Fig.~\ref{fig:fitparams} we present the relation between the best-fit values of the posteriors for the parameters of all NFW and Einasto fits and we give the Pearson correlation coefficient for each pair of parameters. We define the best-fit parameters from the maximum point of multi-dimensional posterior distributions. In the left panel, we show with big stars the mean concentration per $M_{200}$ bin for NFW and compare to the concentration-mass relations in \citet{cui2018}. The concentration-mass relation obtained in this work with NFW follows the flat relation of the \texttt{GADGET-X} clusters results from \textsc{The Three Hundred} in \citet{cui2018}. See Sect.~\ref{sec:massreconstruction} for details. As a reference, we also show the horizontal $c_{200} = 3$ line that corresponds to the concentration value fixed in \citet{giocoli2023} to obtain less biased masses when fitting tNFW density models.
\FloatBarrier

\subsection{Reconstructed mass profiles}
\begin{figure*}[h]
  \centering
  \includegraphics[trim={0pt 0pt 0pt 0pt},scale=0.38]{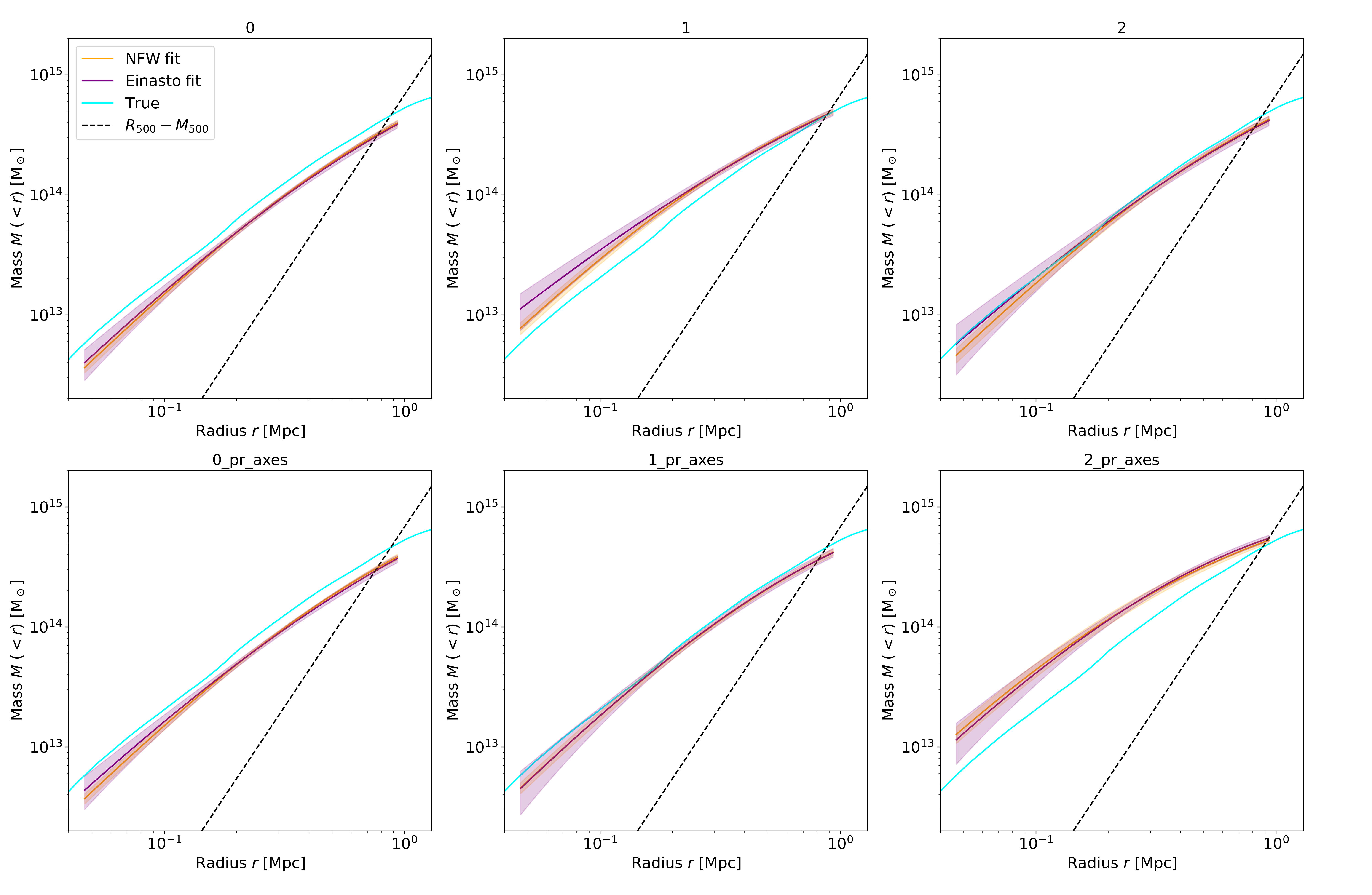}
  \caption{Mass profiles reconstructed from the convergence map fits for the 0306 cluster in the snapshot 101 ($z=0.817$). The orange and purple profiles correspond to NFW and Einasto models, respectively. We give the mean profiles with $1\sigma$ contours. The cyan profile is the spherical mass profile obtained from the simulation. The black dashed line shows the $R_{500}-M_{500}$ relation. }
  \label{fig:massfit}
\end{figure*}

In Fig.~\ref{fig:massfit} we show the mass profiles corresponding to the projected mass density fits of Fig.~\ref{fig:sigmafit} with NFW and Einasto models for the 0306 cluster in the snapshot 101. We present the profiles for the projections along the 3 main axes of the simulation on top, and in the bottom the profiles for the main inertia moment axes of the cluster. The dashed line shows the $R_{500} - M_{500}$ relation given by the definition: $M_{500} /\frac{4}{3}\pi R_{500}^3 = 500 \rho_{\mathrm{crit}}$. The cyan profile shows the spherical true mass profile (Sect.~\ref{sec:icmprofiles}). The departure of the NFW and Einasto profiles from the real spherical mass profile at $\sim R_{500}$ will determine the difference between the true $M_{500}$ and the NFW and Einasto estimates. 
\FloatBarrier

\section{Mass bias scatter distributions}
\begin{figure*}
        \centering
        \begin{minipage}[b]{0.48\textwidth}
        \includegraphics[trim={0pt 0pt 0pt 0pt},scale=0.35]{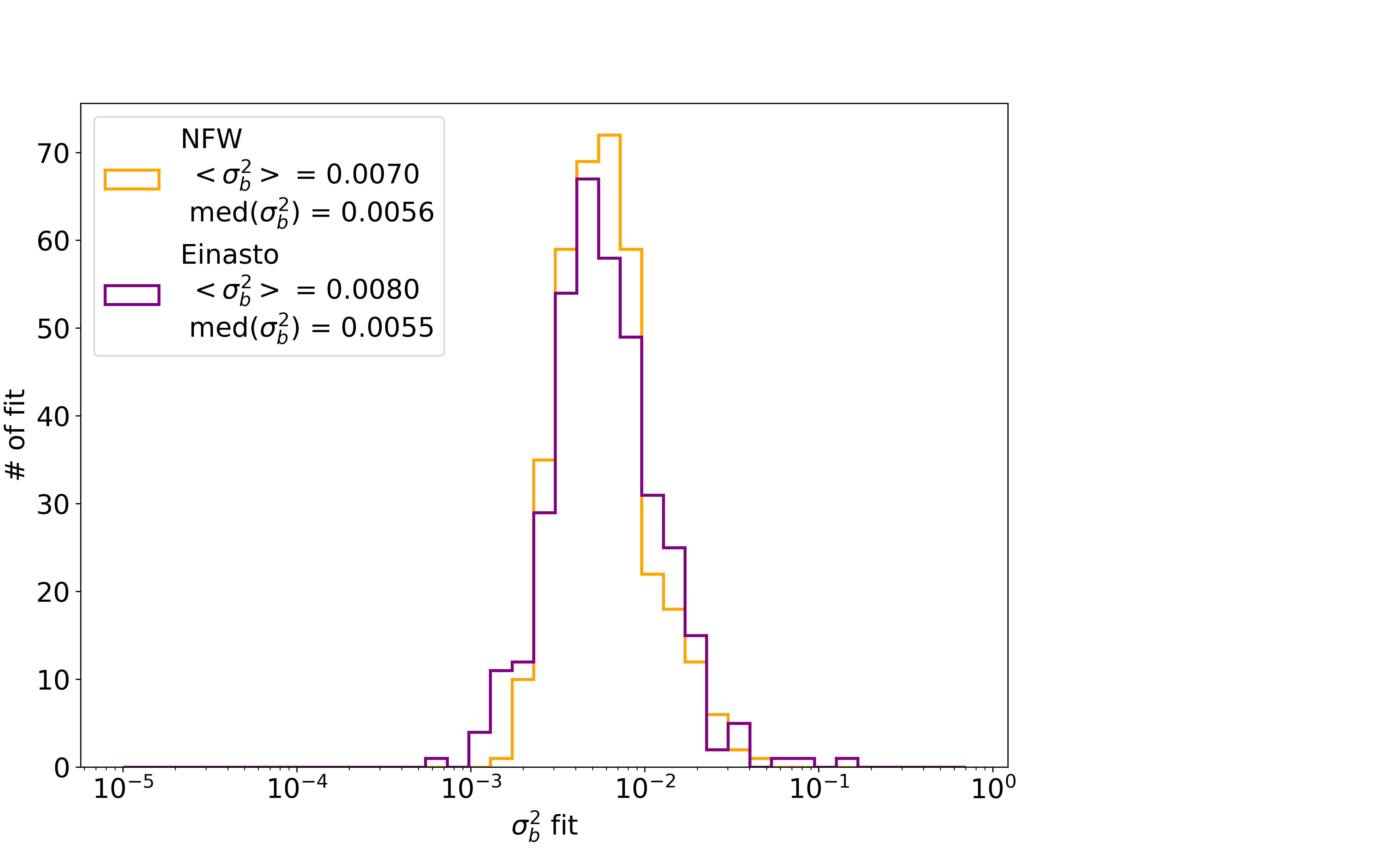}
        \end{minipage}
        \hfill
        \begin{minipage}[b]{0.48\textwidth}
         \includegraphics[trim={0pt 0pt 0pt 0pt},scale=0.35]{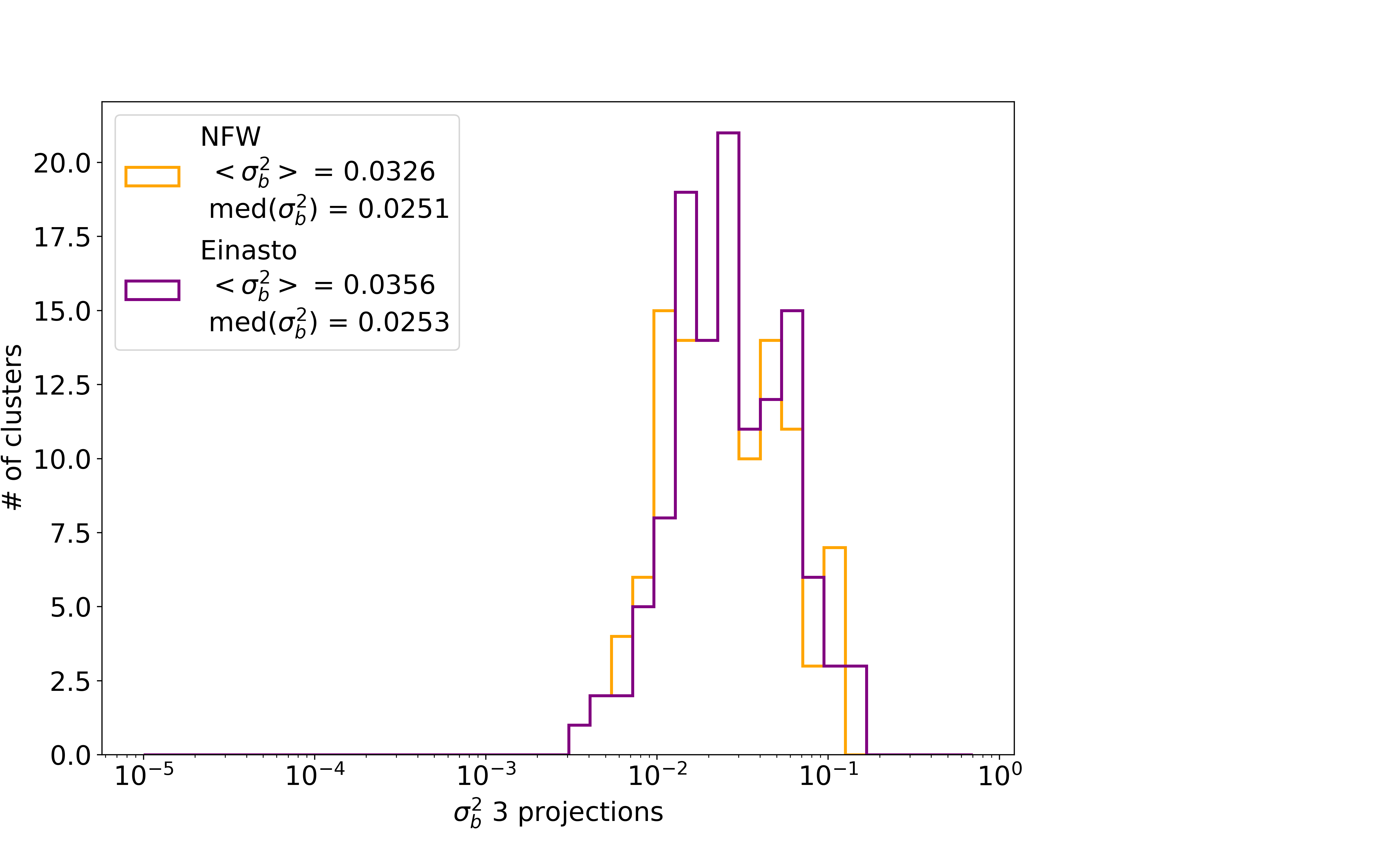}
        \end{minipage}
        \caption{Variance of mass bias for all clusters. Left: scatter induced by uncertainties on the fit, for all the projections of all the clusters in our sample. Right: variance of the bias in the mass reconstruction of each cluster.}
  \label{fig:scattersboth}
\end{figure*}

We present in Fig.~\ref{fig:scattersboth} the distributions of $\sigma^2_{b,\mathrm{fit}}$ (left) and $\sigma^2_{b,\mathrm{3 \; projections}}$ (right) for all the clusters in our sample. The histograms correspond to the bias variance for the NFW (orange) and Einasto (purple) models.
\FloatBarrier

\section{Validation of the projection effect}
\label{sec:validation100}

As we are interested in checking the real scatter due to the projection effects, we produced for 4 different clusters 100 $\kappa\text{-maps}$ along random projections. We chose one cluster per snapshot, with diverse dynamical states, mostly unrelaxed \citep[Table~\ref{tab:100projs}, dynamical state indicators defined following the definitions in ][]{deluca2021}.
    
\renewcommand{\arraystretch}{1.4}
\tiny
    \begin{table*}[]
      \centering
      \caption{Dynamical state indicators at $R_{200}$ and $R_{500}$ for the clusters analysed from 100 projections.}
      \tiny
        \begin{tabular}{c|c|c|c|c|c|c|c|c}
          \hline
          \hline
             Snapshot & $z$ & Cluster & $f_s (R_{200})$ & $\Delta_r (R_{200})$ & Dynamical state within $R_{200}$ &  $f_s (R_{500})$ & $\Delta_r (R_{500})$ & Dynamical state within $R_{500}$\\ \hline
             101 & 0.817 & 0306 & 0.19 & 0.11 & Disturbed &  0.09 & 0.10 & Intermediate \\
             104 & 0.700 & 0206 & 0.18 & 0.08 & Intermediate & 0.16 & 0.16 & Disturbed\\
             107 & 0.592 & 0046 & 0.12 & 0.03 & Intermediate & 0.07 & 0.00 & Relaxed \\
             110 & 0.490 & 0198 & 0.18 & 0.14 & Disturbed & 0.19 & 0.15 & Disturbed \\\hline
        \end{tabular}
        \vspace*{0.2cm}  
       
        \label{tab:100projs}
    \end{table*}
\normalsize    
\FloatBarrier

    For each map we followed the NFW and Einasto mass reconstruction procedure described in Sect.~\ref{sec:massreconstruction}. Therefore, for the considered four clusters, we measured the $\kappa$-bias of each projection. The mean bias and variance for the 100 projections, for the 3 main axes and for the 3 inertia moment axes are summarised in Table~\ref{tab:100bias}. Although considering the principal axes of the clusters regarding their moments of inertia is a way to account for the whole dispersion due to projection, it gives overestimated scatter values with respect to the dispersion obtained from 100 projections. The dispersion, when estimated with only 3 data points, can be very unreliable, but for the 3 inertia moment axes it tends to be overestimated, while for the 3 main (that is, random) axes it can be both over and underestimated. Hence, for the analysis with the full sample of clusters we decided to remove the 0\_pr\_axes, 1\_pr\_axes, and 2\_pr\_axes projections.  
    
    \renewcommand{\arraystretch}{1.4}
    \tiny
    \begin{table*}[]
      \centering
      \caption{Mean bias and variance due to projection effect for the 4 clusters studied with 100 maps. }
      \tiny

        \begin{tabular}{c|c|c|c|c|c|c|c|c|c}
          \hline
          \hline
             Density model & Snapshot & $z$ & Cluster & $< b^{\kappa}>$&  $\sigma_{b, \mathrm{100 \; projections}}^2$ &   $< b^{\kappa}>$& $\sigma_{b, \mathrm{3 \; projections}}^2$&  $< b^{\kappa}>$& $\sigma_{b, \mathrm{3 \; projections}}^2$ \\
             ~ & ~& ~& ~&  100 random & 100 random & 3 main & 3 main & 3 inertia  & 3 inertia \\ \hline
             NFW \\ \hline
             ~ & 101 & 0.817 & 0306 & 0.1155 & 0.0229 & 0.1490 & 0.0219 & 0.1314 & 0.0426 \\
             ~ & 104 & 0.700 & 0206 &  0.1233 & 0.0453 & 0.2036 & 0.0151 & 0.0671 & 0.1442 \\
             ~ & 107 & 0.592 & 0046 &  -0.0450 & 0.0212 & -0.0317 & 0.0176 & -0.0344 & 0.0350 \\
             ~ & 110 & 0.490 & 0198 &  0.0484 & 0.0355 & 0.0231 & 0.0495 & 0.1161 & 0.0281\\\hline
             Einasto  \\ \hline
             ~ & 101 & 0.817 & 0306 &  0.1047 & 0.0359 & 0.1712 & 0.0266 & 0.1259 & 0.0613 \\
             ~ & 104 & 0.700 & 0206 &  0.1312 & 0.0500 & 0.2162 & 0.0158 & 0.0735 & 0.1527 \\
             ~ & 107 & 0.592 & 0046 &  -0.0618 & 0.0212 & -0.0545 & 0.0184 & -0.0447 & 0.0301 \\
             ~ & 110 & 0.490 & 0198 &  0.0716 & 0.0387 & 0.0771 & 0.0378 & 0.1393 & 0.0373\\\hline
        \end{tabular}
        \vspace*{0.2cm}
    \begin{tablenotes}
    \small
    \centering
    \item \textbf{Notes.} The three main projections are the so-called 0, 1, and 2 or $x, y$, and $z$ projections. The 3 intertia projections are: 0\_pr\_axes, 1\_pr\_axes, and 2\_pr\_axes.
    \end{tablenotes}
    \vspace*{0.2cm}
    
    \label{tab:100bias}
        
    \end{table*}
\normalsize     
\FloatBarrier

\section{Scatter from spherical mass estimates and from projected masses within a fixed aperture}
\label{sec:cyl_vs_sph}

In this section we demonstrate that masses cylindrically integrated within a fixed aperture are less scattered than spherically integrated masses. We use simulated mass density profiles of different shapes following the NFW (Eq.~\ref{eq:nfw}), gNFW, tNFW, and Hernquist \citep{hernquist} models.

We define the gNFW profile as       
\begin{equation}
    \rho_{\rm gNFW}(r) = \frac{\rho_s}{\left(\frac{r}{r_s}\right)^{\gamma}\left( 1  + \left(\frac{r}{r_s}\right)^{\alpha}\right)^{\frac{\beta-\gamma}{\alpha}}},
    \label{eq:gnfw}
\end{equation}
with $\alpha =3$, $r_s=R_{\Delta}/c_{\Delta}$ and:
\begin{equation}
     \rho_s = \frac{c_{\Delta}^3 \Delta\rho_{\mathrm{crit}} (-3+\gamma)}{-3(c_{\Delta}^3)^{1-\gamma/3} \;  _2 F_1((\beta-\gamma)/3, 1-\gamma/3, 2-\gamma/3, -c_{\Delta}^3)}.
\end{equation}
Here $_2 F_1$ is a Gaussian hypergeometric function. The tNFW is given by
\begin{equation}
  \rho_{\rm tNFW}(r) = \frac{\rho_{s}}{r/r_s(1+r/r_s)^2} \left( \frac{\tau^2}{\tau^2+(r/r_s)^2}\right)^{\eta},
  \label{eq:tnfw}
\end{equation}
with $\eta =1$,  $r_s=R_{\Delta}/c_{\Delta}$ and:
\begin{equation}
     \rho_s  = \frac{c_{\Delta}^3 \Delta \rho_{\mathrm{crit}}}{3I(c_{\Delta})},
\end{equation}
where,
\begin{equation}
  \begin{split}
    I(c_{\Delta}) =  \tau^{2}\frac{-1 \left[2(1+ \tau^2) - (-1+ \tau^2)\ln( \tau^2)\right]}{2(1+ \tau^2)^2} + \\
        \tau^{2}\frac{\frac{2(1+ \tau^2)}{(1+c_{\Delta})} +   4   \tau {\rm arctan}(c_{\Delta}/ \tau)}{2(1+ \tau^2)^2} + \\
        \tau^{2}\frac{(-1+ \tau^2)\left[2\ln(1+c_{\Delta}) - \ln(c_{\Delta}^2 + \tau^2)\right]}{2(1+ \tau^2)^2}.
   \end{split}
\end{equation}
The Hernquist mass density profile is defined as
\begin{equation}
    \rho_{\rm Hernquist}(r) = \frac{\rho_s}{\frac{r}{r_s}\left( 1  + \frac{r}{r_s} \right)^3},
    \label{eq:hernquist}
\end{equation}
with $r_s=R_{\Delta}/c_{\Delta}$ and:
\begin{equation}
     \rho_s  = \frac{c_{\Delta} \Delta \rho_{\mathrm{crit}}2(1+c_{\Delta})^2}{3}.
\end{equation}

\begin{figure*}[h]
        \centering
        \begin{minipage}{0.49\textwidth}
        \includegraphics[trim={0pt 0pt 0pt 0pt},scale=0.35]{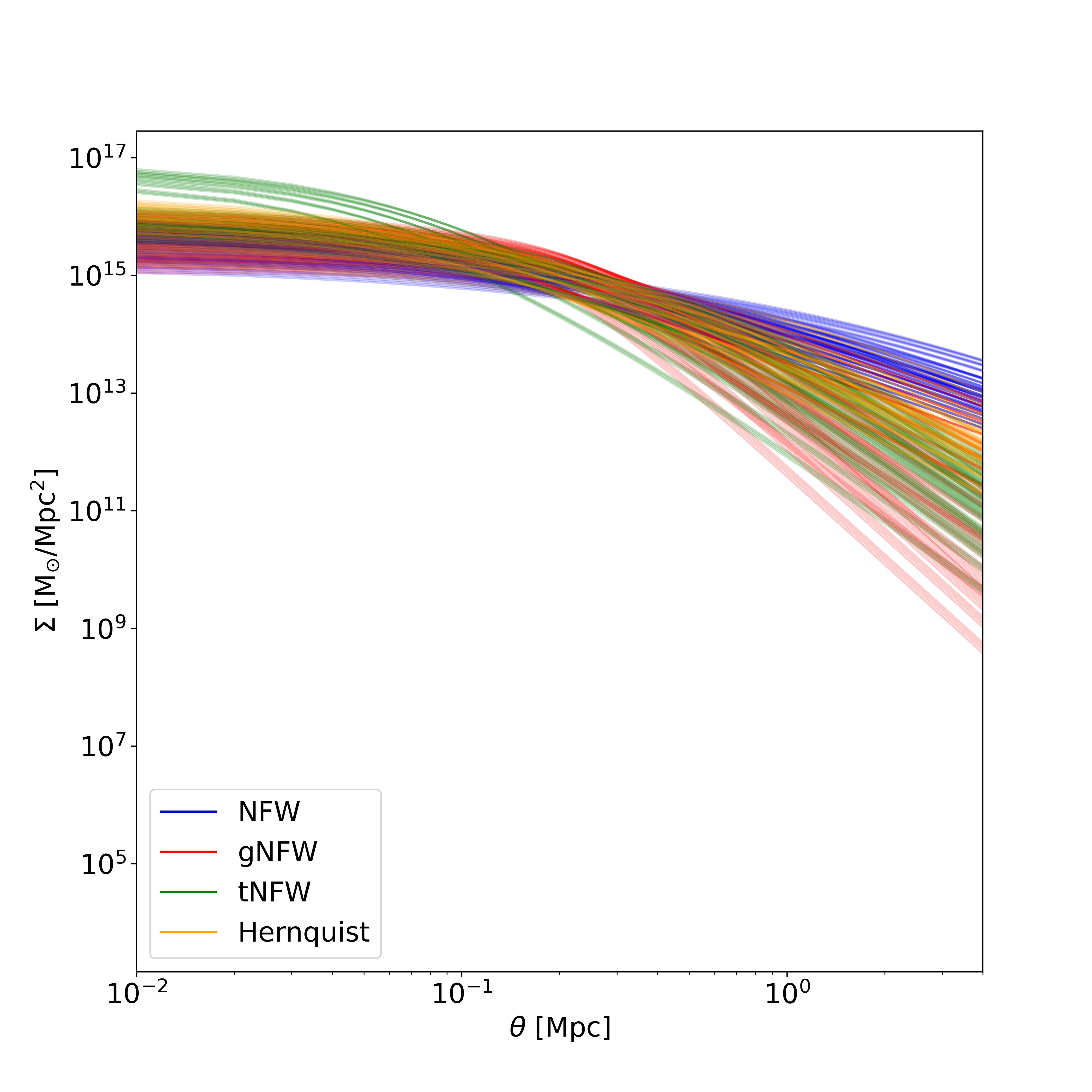}
        \end{minipage}
        \hfill
        \begin{minipage}{0.49\textwidth}
        \includegraphics[trim={0pt 0pt 0pt 0pt},scale=0.37]{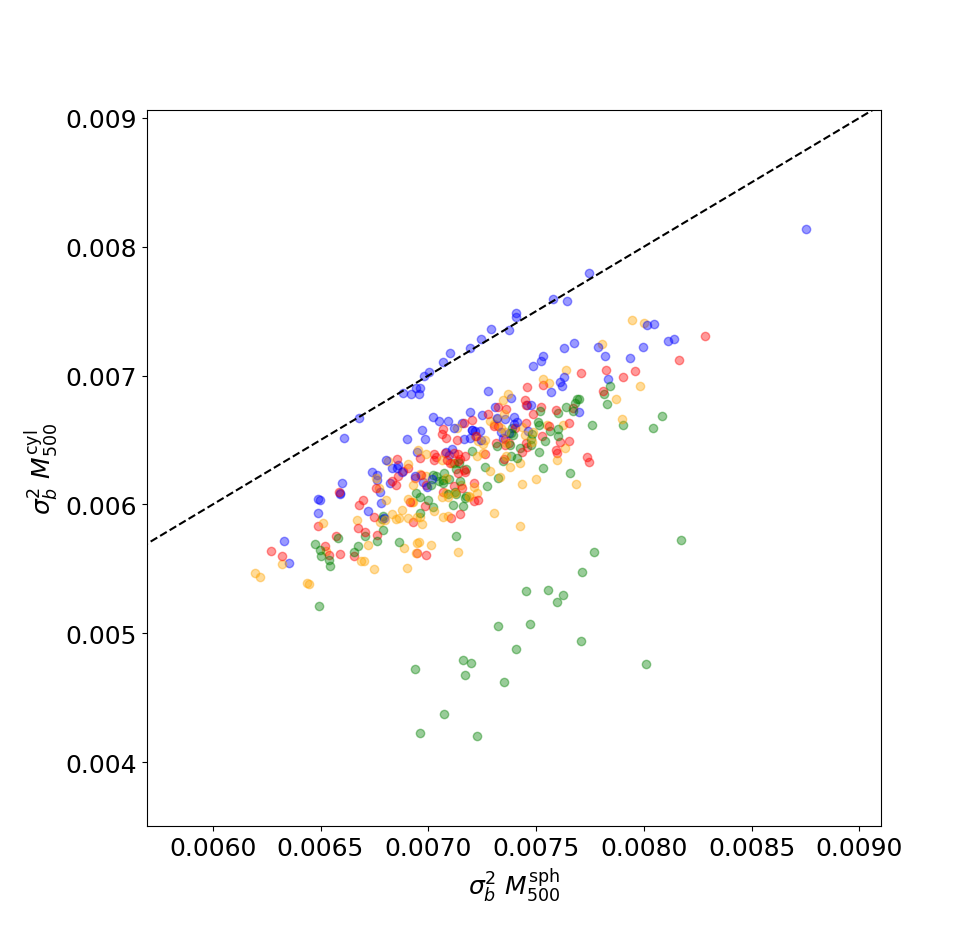}
        \end{minipage}
        \caption{Projected mass density profiles used to quantify the difference in scatter for spherically and cylindrically integrated masses (left). The variance of cylindrically integrated masses with respect to spherically integrated masses (right). The black dashed line corresponds to $\sigma_{b^{\text{sph}}}^{2} = \sigma_{b^{\text{cyl}}}^{2}$. In blue NFW, in red gNFW, in green tNFW, and in orange Hernquist models. }
  \label{fig:dif_models}       
\end{figure*}

We produce profiles with a wide variety of parameters for 5 different initial $M_{500}$ values between $3 \times 10^{14}$~M$_{\odot}$ and $11 \times 10^{14}$~M$_{\odot}$ and for the four different redshifts in our twin samples ($z = [0.490, 0.592, 0.700, 0.817]$). For NFW and Hernquist models we vary $c_{500}$ from 1 to 6. For gNFW and tNFW we fix $c_{500}=4$ and vary $\beta = 3-6$ and $\gamma =0.4 -1$ for gNFW and $\tau = 0.2 -4$ for tNFW. All the considered projected mass density profiles are shown in the left panel in Fig.~\ref{fig:dif_models} to demonstrate the variety of the analysed density shapes.
\FloatBarrier

To mimic a departure from the true mass, for each profile with $M_{500}$ we estimate a distribution of new profiles for masses that vary from $-18 \%$ to $18 \%$ from the true $M_{500}$, $M_{500}^{\text{new sph}}$ (based on an average $\sigma^2_{b, \mathrm{3 \; projections}}$ that we measured from data). For each of the new profiles we can integrate the projected density up to the original $\theta_{500}$,  $M_{500}^{\text{new  cyl}}$. Comparing the new $M_{500}^{\text{new sph}}$ and $M_{500}^{\text{new  cyl}}$ to the original $M_{500}$ we get the spherical and cylindrical biases. From the distribution of biased masses, we take randomly trios to simulate the 3 random projections and estimate their variance: $\sigma_{b^{\text{sph}}}^{2}$ is the variance of three biased spherical masses and with their corresponding cylindrically integrated masses we get $\sigma_{b^{\text{cyl}}}^{2}$.

In the right panel in Fig.~\ref{fig:dif_models} we present the relation between the mean $\sigma_{b^{\text{sph}}}^{2}$ and $\sigma_{b^{\text{cyl}}}^{2}$ for each mass, redshift, and profile shape considered. Same colours as in Fig.~\ref{fig:dif_models} are used to make reference to each density profile type. As expected, the figure shows that the dispersion of cylindrically integrated masses at a given aperture tends to be smaller ($\sigma_{b^{\text{sph}}}^{2} - \sigma_{b^{\text{cyl}}}^{2} \sim  0.001$ to $0.003$) than spherical integrations up to the corresponding radius in each case. The blue points close to the one-to-one line correspond to NFW models with a concentration of $c_{500}=1$.

\section{Building the $Y_{500}-M_{500}$ scaling relation}
\label{sec:appendixD}

We construct our own scaling relation following the parametrisation in Eq.~16 in \citet{arnaud10}:
\begin{equation}
  h(z)^{-2/3}\; Y_{500} = \beta \left[ \frac{M_{500}}{3\times 10^{14}\; \mathrm{h}_{70}^{-1}\; \mathrm{M}_{\odot}}\right]^{\alpha} \mathrm{h}_{70}^{-1},
  \label{eq:SR}
\end{equation}
where $ Y_{500}$ is in Mpc$^2$ and $M_{500}$ in $\mathrm{M}_{\odot}$. Here $h(z)$ is the ratio of the Hubble constant at redshift $z$ to its present value, and, h$_{70}$ the Hubble constant in units of 70 km/s/Mpc. 

We compare two scaling relations that relate the cylindrical or projected $Y_{500}$ to the spherical and cylindrical $M_{500}$ mass, respectively. The spherical quantities are computed from the ICM profiles: $M^{\mathrm{sph}}_{500}$ is the mass profile evaluated at $R_{500}$, that is, the true $M_{500}$ throughout this paper. The cylindrical quantities $M^{\mathrm{cyl}}_{500}$ and $Y^{\mathrm{cyl}}_{500}$ are obtained by integrating the signal within $\theta_{500}$ in the $\kappa\text{-}$ and $y\text{-maps}$. 

\begin{figure*}
        \centering
        \begin{minipage}[b]{0.48\textwidth}
        \includegraphics[trim={0pt 0pt 0pt 0pt},scale=0.35]{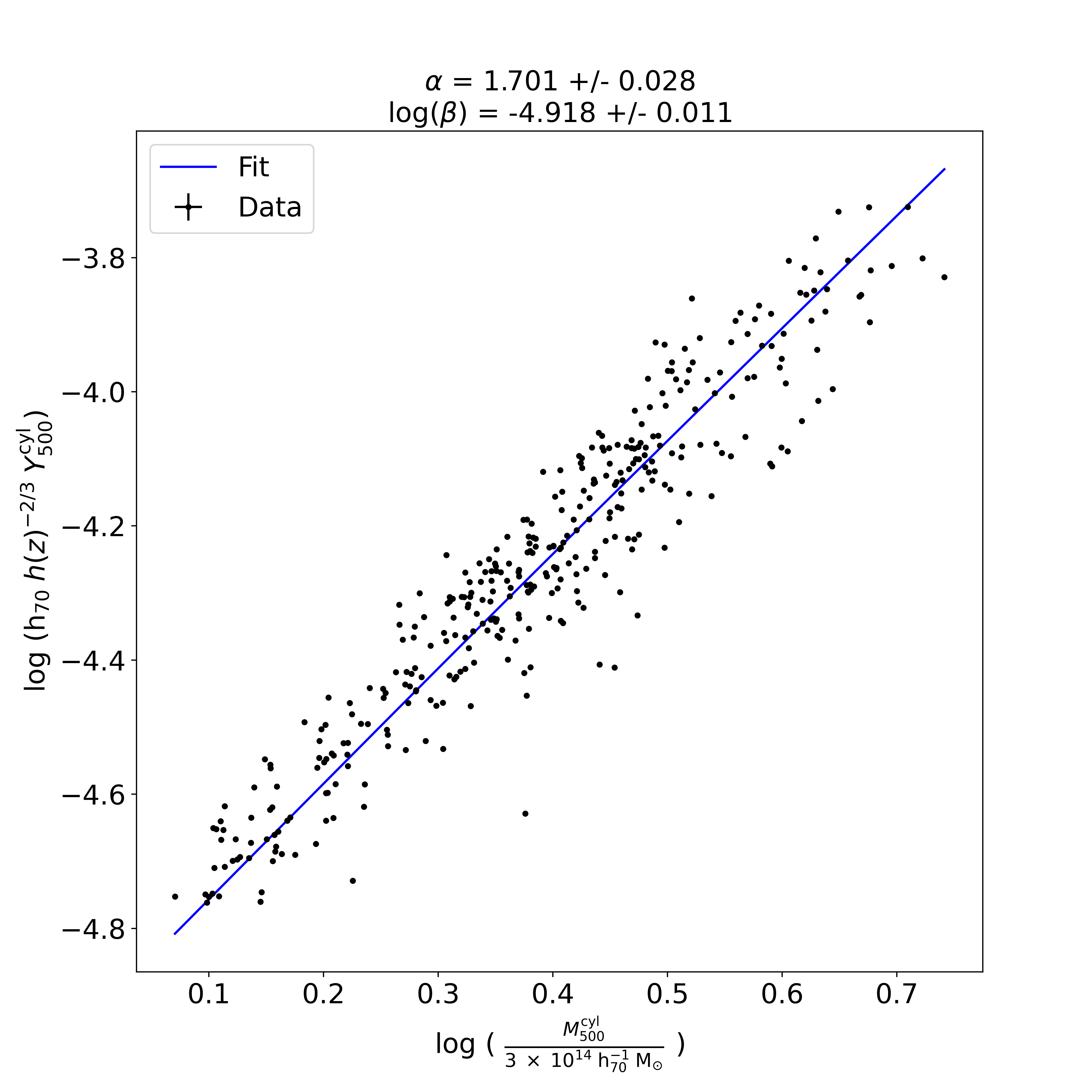}
        \end{minipage}
        \hfill
        \begin{minipage}[b]{0.48\textwidth}
         \includegraphics[trim={0pt 0pt 0pt 0pt},scale=0.35]{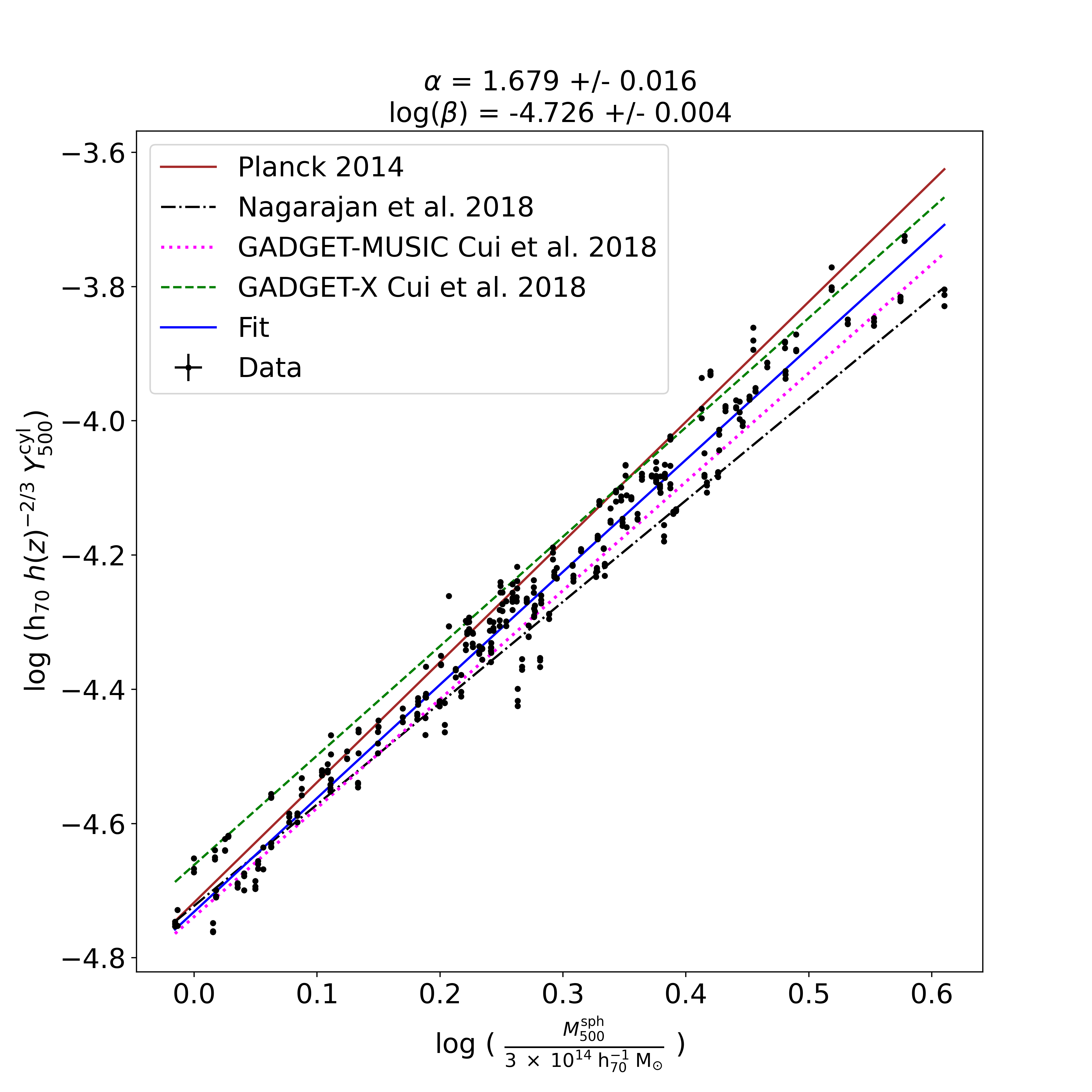}
        \end{minipage}
        \caption{Scaling relations between $Y_{500}$ and $M_{500}$ for cylindrically integrated SZ signal with respect to spherically (right) and cylindrically (left) integrated masses. Black dots correspond to the values per cluster in our sample, while the blue line shows the best-fit relation. The solid brown line shows the result in \cite{planck2014a}, the black dash-dotted is from \cite{Nagarajan_2018}. The pink dotted and green dashed relations are the results in \cite{cui2018} for the \texttt{GADGET-MUSIC} and \texttt{GADGET-X} simulations. }
  \label{fig:SR}
\end{figure*}

In each case, we perform a fit of Eq.~\ref{eq:SR} using the orthogonal BCES method \citep{bces}. Uncertainties on fit parameters are estimated from 100000 bootstrap resamples of the data. We show the scaling relations in Fig.~\ref{fig:SR}, as well as the best-fit parameters (top legend in the figure). The slopes of both relations are consistent, but since the cylindrically integrated masses are larger than spherically integrated ones, the intercepts are not compatible.

We compare, as in Fig.~10 in \cite{cui2018}, the $Y^{\mathrm{cyl}}_{500} - M^{\mathrm{sph}}_{500}$ relation to results in the literature. In \cite{cui2018}, the authors computed the scaling relation using all \textsc{The Three Hundred} clusters at $z=0$, for both \texttt{GADGET-X} and \texttt{GADGET-MUSIC} flavours. The scaling relation obtained with our sample has a lower normalisation and slightly steeper slope than the results obtained with \texttt{GADGET-X} simulations in \cite{cui2018}. The differences may be due to both the considered cluster sample and the redshift range. The \texttt{GADGET-MUSIC} scaling relation from \cite{cui2018} has an even smaller intercept value, showing the impact of the hydrodynamical model on the simulation. 
In the same figure we also compare the results to the scaling relations obtained from observations in \cite{planck2014a} and \cite{Nagarajan_2018}. Our scaling relation lies between both observational results, with a flatter slope than \cite{planck2014a} and steeper than \cite{Nagarajan_2018}. Given the offset between spherical $M^{\mathrm{sph}}_{500}$ and cylindrical  $M^{\mathrm{cyl}}_{500}$ masses, the observational results shown in the right panel in Fig.~\ref{fig:SR} are shifted with respect to the scaling relation in the left panel and are not comparable.

\FloatBarrier

\end{appendix}
\end{document}